\documentclass[twocolumn]{aastex}
\usepackage{amsmath,amsfonts,amssymb,amscd,amsthm}
\usepackage{xcolor}

\renewcommand{\vec}[1]{{\bf #1}}
\renewcommand{\hat}[1]{{\bf #1}}

\begin{document}

\title{On the Turbulent Reduction of Drifts for Solar Energetic Particles}









\author{J.P. van den Berg\altaffilmark{1}\altaffilmark{2}, N.E. Engelbrecht\altaffilmark{1}, N. Wijsen\altaffilmark{3}, R.D. Strauss\altaffilmark{1}}

\altaffiltext{1}{Center for Space Research, North-West University, Potchefstroom, 2522, South Africa; 24182869@nwu.ac.za}
\altaffiltext{2}{South African National Space Agency, Hermanus, 7200, South Africa}
\altaffiltext{3}{Department of Mathematics/Centre for mathematical Plasma Astrophysics, KU Leuven, Belgium}

\begin{abstract}
Particle drifts perpendicular to the background magnetic field have been proposed by some authors as an explanation for the very efficient perpendicular transport of solar energetic particles (SEPs). This process, however, competes with perpendicular diffusion caused by magnetic turbulence, which can also disrupt the drift patterns and reduce the efficiency of drift effects. The latter phenomenon is well known in cosmic ray studies, but not yet considered in SEP models. Additionally, SEP models which do not include drifts, especially for electrons, use turbulent drift reduction as a justification of this omission, without critically evaluating or testing this assumption. This article presents the first theoretical step for a theory of drift suppression in SEP transport. This is done by deriving the turbulence-dependent drift reduction function with a pitch-angle dependence, as applicable for anisotropic particle distributions, and by investigating to what extent drifts will be reduced in the inner heliosphere for realistic turbulence conditions and different pitch-angle dependencies of the perpendicular diffusion coefficient. The influence of the derived turbulent drift reduction factors on the transport of SEPs are tested, using a state-of-the-art SEP transport code, for several expressions of theoretically-derived perpendicular diffusion coefficients. It is found, for realistic turbulence conditions in the inner heliosphere, that cross-field diffusion will have the largest influence on the perpendicular transport of SEPs, as opposed to particle drifts.
\end{abstract}



\section{Introduction}
\label{sec:Intro}

The perpendicular transport of solar energetic particles (SEPs) has received more attention in recent years in an attempt to explain observed widespread events \citep[e.g.][]{cliveretal1995, dresingetal2014, gomezherreroetal2015}. The two mechanisms responsible for transport perpendicular to the background magnetic field are perpendicular diffusion and drifts. The latter mechanism has received less attention in the context of SEPs and our knowledge, therefore, is still lacking on this subject. One reason for its neglect might be due to the lack of clear observational evidence of SEPs experiencing drifts. As such, most work on the drifts experienced by SEPs is based on single particle/full-orbit simulations \cite[e.g.][]{kellyetal2012, marshetal2013, battarbeeetal2017, battarbeeetal2018a}, with some theoretical investigations into the drift velocity \citep[e.g. ][]{burnshalpern1968, dallaetal2013} and transport modelling \citep{wijsenetal2020}. A summary of this can be found in \citet{vandenbergetal2020}, with the aspects most relevant to the present study to be highlighted below.

As already mentioned, observational evidence pointing to SEPs experiencing drifts is limited. Observations of the STEREO mission between October 2006 and December 2013 reported by \citet{richardsonetal2014} seem to indicate that solar energetic proton and electron events behave similarly. This follows from comparing the onset or peak delay and the angular position of the peak intensity as a function of the angular separation between the flare and the spacecraft's magnetic footpoint. If drifts were primarily responsible for perpendicular transport, then it would be expected that the electrons and protons should behave differently, although the data are not presented according to the polarity of the magnetic field. The time span under consideration did fall mainly in an $A < 0$ polarity cycle ($A$ is the HMF polarity, defined as having $A = +1$ when the magnetic field points outwards in the Northern hemisphere), but the polarity seen by an observer depends on their position relative to the heliospheric current sheet (HCS). Furthermore, the considered electrons and protons have different rigidities, which complicates direct comparison. More recently, \citet{augustoetal2019} postulated that SEP drifts along the HCS can help SEPs observed in ground level enhancements to reach Earth from a magnetically unconnected event, if the flare or coronal mass ejection and Earth are both close to the HCS. This is based on an analysis of the 2017 September 10 event, when both the active region and Earth were close to the HCS, but the transport time scales (particle streaming along magnetic field to reach $1$~${\rm au}$ and perpendicular transport either through drifts or diffusion) were not calculated to be compared to the time the particles had to travel to Earth (i.e. the time between flare onset and \textit{in situ} onset of SEPs).

Nonetheless, several studies have investigated the effect of drifts on SEPs. \citet{burnshalpern1968} derived expressions for the drift velocities of SEPs in a \citet{parker1958} heliospheric magnetic field (HMF) and concluded that the electric field drift is the dominant process, which shows that particles follow the field line on which they started when this term is transformed into the reference frame co-rotating with the Sun. \citet{dallaetal2013} re-derived these expressions and investigated the other drift terms as well, while \citet{marshetal2013} verified these general predictions by integrating the Newton-Lorentz equation in a uni-polar Parker HMF. These authors found that drifts will be the most pronounced for high energy particles ($\sim 100$~${\rm MeV}$ protons) or partially-ionised heavy ions. This investigation was extended by \citet{battarbeeetal2017} to include neutral sheet drifts in a flat HCS, where it was found that SEP drift patterns will be similar to galactic cosmic ray (CR) drift patterns. Particles can be confined to the HCS in the $qA > 0$ polarity cycle (here $q$ is the particle charge) and might have difficulty reaching an observer on the other side of the HCS, while the energy dependence of drift velocities will cause different observers to measure different spectra. A wavy HCS was considered by \citet{battarbeeetal2018a} and it was found that the HCS can efficiently transport particles to the poles where they will drift more efficiently if they can escape the HCS through scattering.

Based on these models, \citet{dallaetal2017a} used drifts to attribute the observed energy dependent charge state of iron to the mass-to-charge-ratio dependence of drifts and not the acceleration process \citep[see e.g.][and references therein]{reames1999, reames2013}. Similarly, \citet{dallaetal2017b} attributed the temporal evolution of the iron-to-oxygen-ratio to drifts and not the rigidity-dependent mean free path (MFP) caused by turbulence generated by streaming protons of similar rigidity \citep[see again][and references therein]{reames1999, reames2013}. \citet{battarbeeetal2018b} used drifts to model the ground level enhancement event of 2012 May 17 with mixed results: a source $80^{\circ}$ wider in longitude than the inferred coronal mass ejection was needed for particles to reach the STEREO spacecraft and the flux at Mercury was overestimated. 

One aspect of the full-orbit simulations, discussed up to now, which is potentially highly unrealistic is the pitch-angle scattering implemented as random adjustments of the velocity vector at Poisson-distributed scattering intervals (governed by the parallel MFP). A particle in these simulations therefore follows an unperturbed motion between scatterings instead of being continuously subjected to turbulent fluctuations perturbing the velocity vector, as one would expect in a realistic scenario involving broadband turbulence \cite[see, e.g,][]{brunocarbone2016}. This would explain why \citet{marshetal2013} found that pitch-angle scattering and the parallel MFP have little effect on drifts, because the particles experience maximum drift between the occasional scatterings. The full-orbit simulations also do not contain perpendicular diffusion, which will be present in turbulence with transverse complexity \cite[see e.g.][]{qinetal2002a, qinetal2002b}, unlike the initial work of \citet{kellyetal2012}, where a random magnetic foot point motion was added to a \citet{parker1958} HMF in order to simulate the random walk of magnetic field lines, but the velocity vector was also adjusted randomly at Poisson-distributed times. In these simulations the particles show a diffusive spread perpendicular to the background HMF with a slight asymmetry due to drifts.

The recent modelling efforts of \citet{wijsenetal2020} used the focused transport equation to investigate drifts in a uni-polar HMF with a pitch-angle independent perpendicular diffusion coefficient. These authors verified, for $3-36$~${\rm MeV}$ protons, that different observers will see different spectra, and found that perpendicular diffusion will diminish the effects of drifts (i.e. break the latitudinal and longitudinal asymmetry with respect to the injection site caused by drifts). Their investigation into an SEP reservoir \citep[different observers measuring the same intensity and spectra during the fairly isotropic decay phase of an SEP event; see][for a summary]{reames2013} formed by a co-rotating interaction region, yielded an interesting result: It is difficult to form a reservoir in the presence of drifts due to the energy dependence of the drift speed and the preferred direction of the drift velocity, causing different energy channels to decay at different rates, different observers to measure different intensities, and the decay phase to show significant anisotropies. \citet{wijsenetal2020} concluded that if drifts are present, then perpendicular diffusion must be strong enough to smear out the effects of drifts in order to see the reservoir effect.

A major shortcoming of all these studies is the neglect of the turbulent reduction of drifts. Drifts of cosmic rays, entering as a second order effect compared to the first order diffusion processes, have been thoroughly investigated and it has been found that drift effects must be suppressed in models to fit observations \cite[see, e.g.,][]{burgeretal2000}. This is due to turbulent fluctuations disrupting the large scale drifts and therefore decreasing the drift velocity \citep[see][for a summary]{engelbrechtetal2017}. Numerical simulations \citep[e.g.][]{minnieetal2007, tautzshalchi2012} have given clues to the circumstances leading to a reduction in drifts: 1) drift reduction do not seem to occur in purely magnetostatic slab turbulence; 2) the drifts also seem to be affected by the diffusion coefficients (probably because different turbulence conditions will yield different diffusion and drift conditions); 3) drifts decrease with an increase in the turbulence strength, due to stronger turbulent drifts and 4) drifts decrease with a decrease in the particle energy, due to low energy particles being perturbed more than high energy particles \citep{burgervisser2010, engelbrechtetal2017}. It is important to note that all studies on this have only considered isotropic particle distributions and do not consider pitch-angle dependencies.
 
It is then the aim of this paper to investigate the possible pitch-angle dependence of the turbulent drift reduction factor and its implications for SEP drifts. The investigation begins in Section~\ref{sec:Reduce} with a summary of Appendix~\ref{apndx:Derivation}, where the turbulence-reduced guiding centre velocity is derived conceptually. This section also gives a summary of what is known about drift suppression in the isotropic case and a derivation of the pitch-angle dependent drift reduction factor. The drift reduction factor will be evaluated for realistic turbulence and scattering conditions in Appendix~\ref{apndx:Predict}, while the effect of this on the transport of solar energetic protons will be presented in Section~\ref{sec:SEPs}. Lastly, the implications of this investigation will be discussed in Section~\ref{sec:Discuss}.


\section{Drift Reductions including a Pitch-angle Dependence}
\label{sec:Reduce}

The derivation of the drift reduction factor with a pitch-angle dependence will be given in this section, together with a discussion of this factor in the isotropic limit and the origin of the drift reduction, based on the derivations in Appendix~\ref{apndx:Derivation}. Appendix~\ref{apndx:Derivation} presents a conceptual derivation \citep[following][]{engelbrechtetal2017} to show that the gyro-phase averaged guiding centre (GC) velocity in the solar wind (SW) frame,
\begin{align}
\label{eq:ParkerDriftVelocity}
\vec{v}_{\rm gc} & = \sqrt{f_s} \, \mu v \, \hat{b}_0 + \vec{v}_{\rm sw} + f_s^{3/2} \frac{\mu p}{q B_0} \, \hat{b}_0 \times [(\hat{b}_0 \cdot \vec{\nabla}) \vec{v}_{\rm sw} \, + \nonumber \\
 & \;\;\;\;\; (\vec{v}_{\rm sw} \cdot \vec{\nabla}) \hat{b}_0] \, + \frac{1 + \mu^2}{4} \frac{v p}{q B_0} (\vec{\nabla} f_s \times \hat{b}_0) \, + \nonumber \\
 & \;\;\;\;\; f_s \frac{v p}{q B_0} \left\lbrace \mu^2 (\vec{\nabla} \times \hat{b}_0)_{\perp} + \frac{1 - \mu^2}{2} \times \right. \nonumber \\
 & \;\;\;\; \left. \left[ (\vec{\nabla} \times \hat{b}_0)_{\parallel} + \frac{\hat{b}_0 \times \vec{\nabla} B_0}{B_0} \right] \right\rbrace ,
\end{align}
is modified by the effects of turbulence through a so-called drift reduction factor, which can be written in terms of the gyrofrequency $\omega_c$ and a perpendicular velocity correlation function's decorrelation time as \citep{biebermatthaeus1997}
\begin{equation}
\label{eq:BAM97}
f_s = \frac{1}{1 + (\omega_c \tau)^{-2}} .
\end{equation}
For Eq.~\ref{eq:ParkerDriftVelocity} it was assumed that the SW velocity $\vec{v}_{\rm sw}$ is constant and radial and that the background magnetic field $\vec{B}_0$ is stationary (see Appendix~\ref{apndx:Derivation} for the full expression in the general case). It was also assumed that the turbulence is weak, since $f_s$ is derived from the velocity correlation of a charged particle in a uniform magnetic field with turbulence causing an exponential decorrelation of the perpendicular velocity on a time scale $\tau$ \citep[following][]{biebermatthaeus1997}, while this form of the GC velocity is calculated using a perturbation of the fields. Here, $\mu$, $v$, $p$, and $q$ is the particle's pitch-cosine, speed, momentum, and charge in the SW frame, respectively, $\hat{b}_0 = \vec{B}_0 / B_0$ is a unit vector along the large scale average/background magnetic field, the subscript $\parallel$ and $\perp$ refer to directions parallel and perpendicular to $\vec{B}_0$, and $\omega_c = |q| B_0 / m$ is the particle's cyclotron frequency, with $m$ the particle's relativistic mass in the SW frame. The terms in Eq.~\ref{eq:ParkerDriftVelocity} describe the streaming of particles along $\hat{b}_0$, advection by the SW, drifts caused by the streaming of particles into different SW conditions and the advection of the magnetic field by the SW, spatially varying turbulence conditions, curvature drift perpendicular and parallel to the magnetic field \citep[the parallel term will be absent if the fields are evaluated at the GC's position and not the particle's position;][]{rossiolbert1970, burgeretal1985, wijsen2020}, and gradient drift.

Notice that both the streaming and drifts in Eq.~\ref{eq:ParkerDriftVelocity} are reduced in the presence of turbulence. \emph{The streaming terms} ($\mu v \, \hat{b}_0 + (1 - \mu^2) v p (\vec{\nabla} \times \hat{b}_0)_{\parallel} / 2 q B_0$) \emph{are reduced because the particle is streaming along the total magnetic field and not just the background magnetic field.} \emph{The perpendicular drift terms are reduced because the smooth drift pattern is disrupted by turbulent drifts.} Indeed, the perturbed part of the velocity (see Appendix~\ref{apndx:Derivation} for details) can be used to calculate the perpendicular diffusion coefficients \citep[see e.g.][]{fraschettijokipii2011, straussetal2016} and the first term in the perturbed component ($\mu v \, \delta \vec{B} / B_0$, where $\delta \vec{B}$ is the fluctuating magnetic field) can be interpreted as particles following random walking field lines. It can therefore be seen that \emph{there is an interplay between drifts and the perpendicular diffusion coefficients: in the absence of turbulence, the particle will experience pure drift $(f_s = 1)$, while this drift velocity is reduced as soon as turbulence is present $(f_s < 1)$, but with the additional effect that the particle will then also experience perpendicular diffusion.}

Initially, parametric forms of the drift suppression factor based on fits to results from numerical test particle simulations of drift coefficients in the presence of simulated magnetic turbulence were presented, but the usefulness of such expressions are limited to the simulation's particular turbulence conditions. \citet{engelbrechtetal2017} derived a form which is able to explain the various simulation results both qualitatively and quantitatively,
\begin{equation}
\label{eq:IsotropicDriftReduction}
f_s = \frac{1}{1 + (\lambda_{\perp}^0 / R_L)^2 (\delta B^2 / B_0^2)} ,
\end{equation}
where $\lambda_{\perp}^0$ is the perpendicular MFP for a nearly isotropic distribution in the absence of focusing, $R_L = v / \omega_c$ is the particle's maximal Larmor radius, and $\delta B^2$ is the total variance of the fluctuations. It is known from simulations that perpendicular diffusion in pure magnetostatic slab turbulence is subdiffusive \citep{qinetal2002a} and hence $\lambda_{\perp}^0 \longrightarrow 0$ with $f_s \longrightarrow 1$. If $\lambda_{\perp}^0 / R_L$ is kept constant, then $f_s \longrightarrow 0$ as the turbulence strength ($\delta B^2 / B_0^2$) increases. Keeping $\delta B^2 / B_0^2$ constant and assuming that $\lambda_{\perp}^0$ is independent of particle speed or only weakly dependent thereon, then $\lambda_{\perp}^0 / R_L \longrightarrow 0$ and $f_s \longrightarrow 1$ for high energy particles \citep{engelbrechtetal2017}. 

Following \citet{engelbrechtetal2017}, it will be assumed that the drift reduction factor is given by Eq.~\ref{eq:BAM97}. \citet{biebermatthaeus1997} argued that $\tau = \ell_c / v$ and introduced a perpendicular decorrelation length $\ell_c = R_L^2 / \kappa_{\rm FL}$, where $\kappa_{\rm FL}$ is the field line random walk (FLRW) diffusion coefficient. \citet{engelbrechtetal2017} relaxed the assumption of decorrelation due to particles following random walking field lines and proposed that $\ell_c = R_L^2 / \lambda_{\perp}^0$, and that $\tau$ is governed by the particle's perpendicular speed $v_{\perp}$ and not $v$. \citet{engelbrechtetal2017} also argued that the turbulence will cause the perpendicular speed to change randomly and therefore interpreted $v_{\perp} \approx v \sqrt{\delta B^2 / B_0^2}$ as a root-mean-squared value.

A choice for $\tau$, however, must now be made which includes pitch-angle information. Following \citet{engelbrechtetal2017}, it will be assumed that the particle's velocity decorrelates by diffusing across the magnetic field, such that
\begin{equation}
\label{eq:PerpDecorlLength}
\ell_c = \frac{r_L^2}{\lambda_{\perp} (\mu)} = \frac{v r_L^2}{3 D_{\perp} (\mu)} ,
\end{equation}
where $r_L$ is the particle's (pitch-angle dependent) Larmor radius and $D_{\perp} (\mu)$ is the pitch-angle dependent perpendicular diffusion coefficient. A pitch-angle dependent perpendicular MFP, $\lambda_{\perp} (\mu) = 3 D_{\perp} (\mu) / v$, has been introduced and related to $D_{\perp} (\mu)$ in such a way that the usual definition of the perpendicular MFP is recovered for an isotropic distribution in the absence of focusing, that is \citep{shalchi2009}
\begin{equation}
\label{eq:perpMFPdef}
\lambda_{\perp}^0 = \frac{1}{2} \int_{-1}^1 \lambda_{\perp} (\mu) d\mu = \frac{3}{2 v} \int_{-1}^1 D_{\perp} (\mu) d\mu .
\end{equation}
This should be a reasonable estimate of the decorrelation time, since the Larmor radius quantifies the extent to which the particle samples fluctuations perpendicular to the background magnetic field and the perpendicular diffusion coefficient quantifies how quickly the particle diffuses perpendicular to the background magnetic field.

Incorporating also a pitch-angle dependence into the root-mean-squared value of the perpendicular speed, $v_{\perp} \approx v \sqrt{1-\mu^2} \sqrt{\delta B^2 / B_0^2}$, the perpendicular decorrelation time is then proposed to be $\tau = \ell_c / v_{\perp} \approx v r_L^2 / 3 D_{\perp} (\mu) v \sqrt{1-\mu^2} \sqrt{\delta B^2 / B_0^2}$. Lastly, the cyclotron frequency can be written in terms of the particle's maximal Larmor radius and speed, i.e. $\omega_c = v / R_L$. With this choice of $\omega_c$ and $\tau$, the product in Eq.~\ref{eq:BAM97} is
\begin{equation*}
\omega_c \tau = \frac{\sqrt{1 - \mu^2}}{D_{\perp} (\mu)} \frac{v R_L}{3} \sqrt{\frac{B_0^2}{\delta B^2}} ,
\end{equation*}
where $r_L = \sqrt{1 - \mu^2} R_L$ was used. Substituting this into Eq.~\ref{eq:BAM97} gives an expression for the drift reduction factor,
\begin{equation}
\label{eq:MuDependentFs}
f_s(\mu) \approx \frac{1}{1 + [D_{\perp} (\mu) / \sqrt{1 - \mu^2} \kappa_A^{\rm ws} ]^2 (\delta B^2 / B_0^2)} ,
\end{equation}
where $\kappa_A^{\rm ws}$ is the isotropic weak-scattering drift coefficient of Eq.~\ref{eq:WeakScatteringLimit}, so that the pitch-angle dependent, turbulence-reduced drift coefficient then becomes
\begin{equation}
\label{eq:MuDependentKA}
D_{xy}(\mu) =  f_s(\mu) \kappa_A^{\rm ws} (\mu) = {\rm sign}(q) f_s(\mu) \frac{1 - \mu^2}{2} v R_L.
\end{equation}
This form of $f_s(\mu)$ will have significant consequences for focused SEPs, as will be explored in the following paragraphs and Appendix~\ref{apndx:Predict}.

Note that Eq.~\ref{eq:MuDependentFs} also encapsulates that which is already known about turbulent drift suppression from computer simulations: 1) $f_s \longrightarrow 1$ as $v$ becomes large (not taking into account the energy dependence of $D_{\perp}$); 2) $f_s \longrightarrow 0$ as $\delta B^2 / B_0^2$ becomes large; 3) if $D_{\perp}$ is consistent with the Shalchi slab hypothesis \citep{shalchi2006}, then $D_{\perp}$ will be zero in pure slab turbulence and hence $f_s = 1$; 4) in most non-linear theories, $D_{\perp}$ is also dependent on the parallel MFP or the pitch-angle diffusion coefficient, such that both parallel and perpendicular diffusion will influence $f_s$. The greatest uncertainty in Eq.~\ref{eq:MuDependentFs} at this point, is the pitch-angle dependence of $D_{\perp} (\mu)$. This is investigated for different theories of the perpendicular diffusion coefficient in the next section and Appendix~\ref{apndx:Predict}.

\begin{figure*}
\begin{center}
\includegraphics[trim=5mm 13mm 14mm 15mm, clip, width=0.5\textwidth]{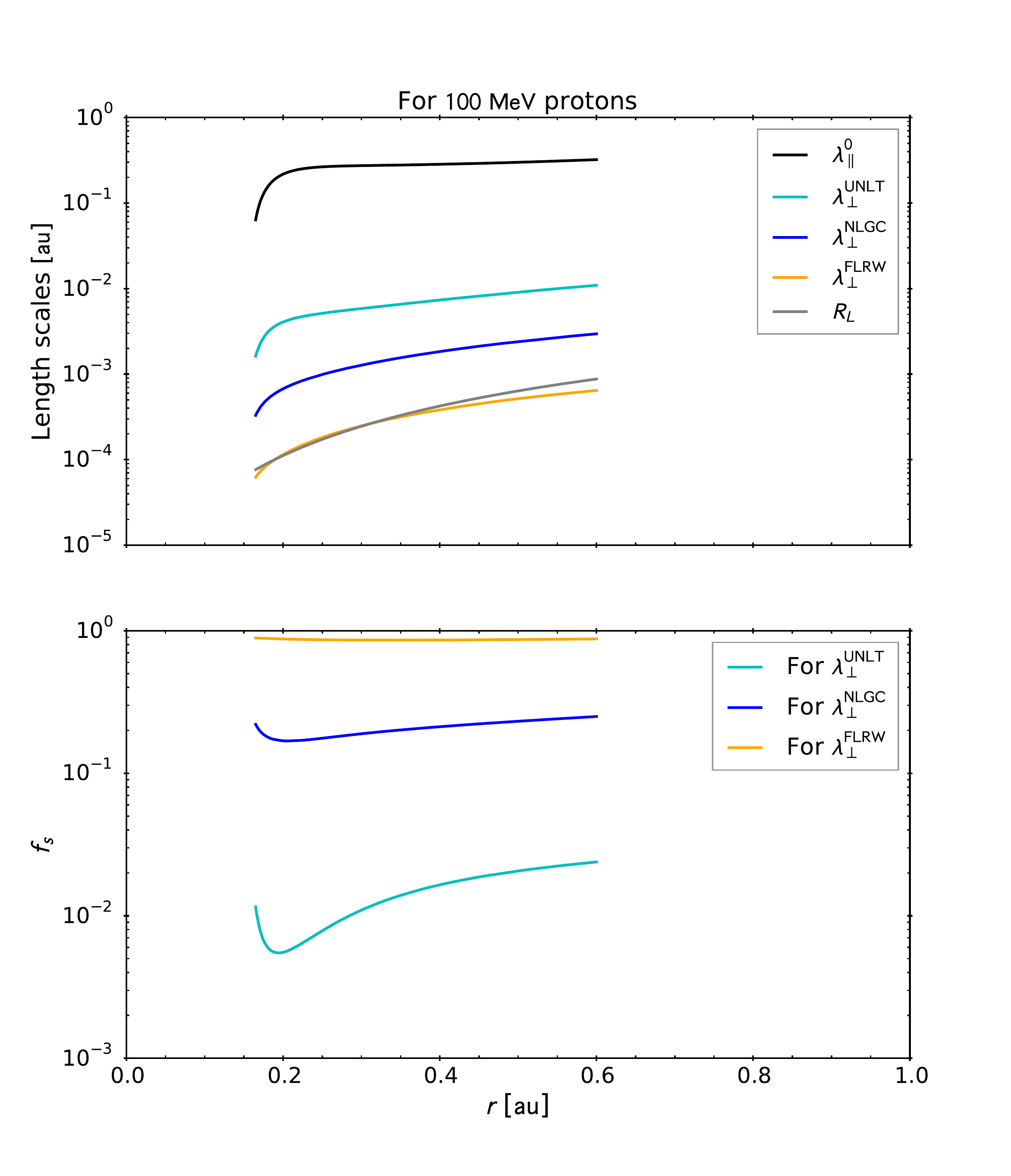}
\includegraphics[trim=15mm 13mm 14mm 15mm, clip, width=0.48\textwidth]{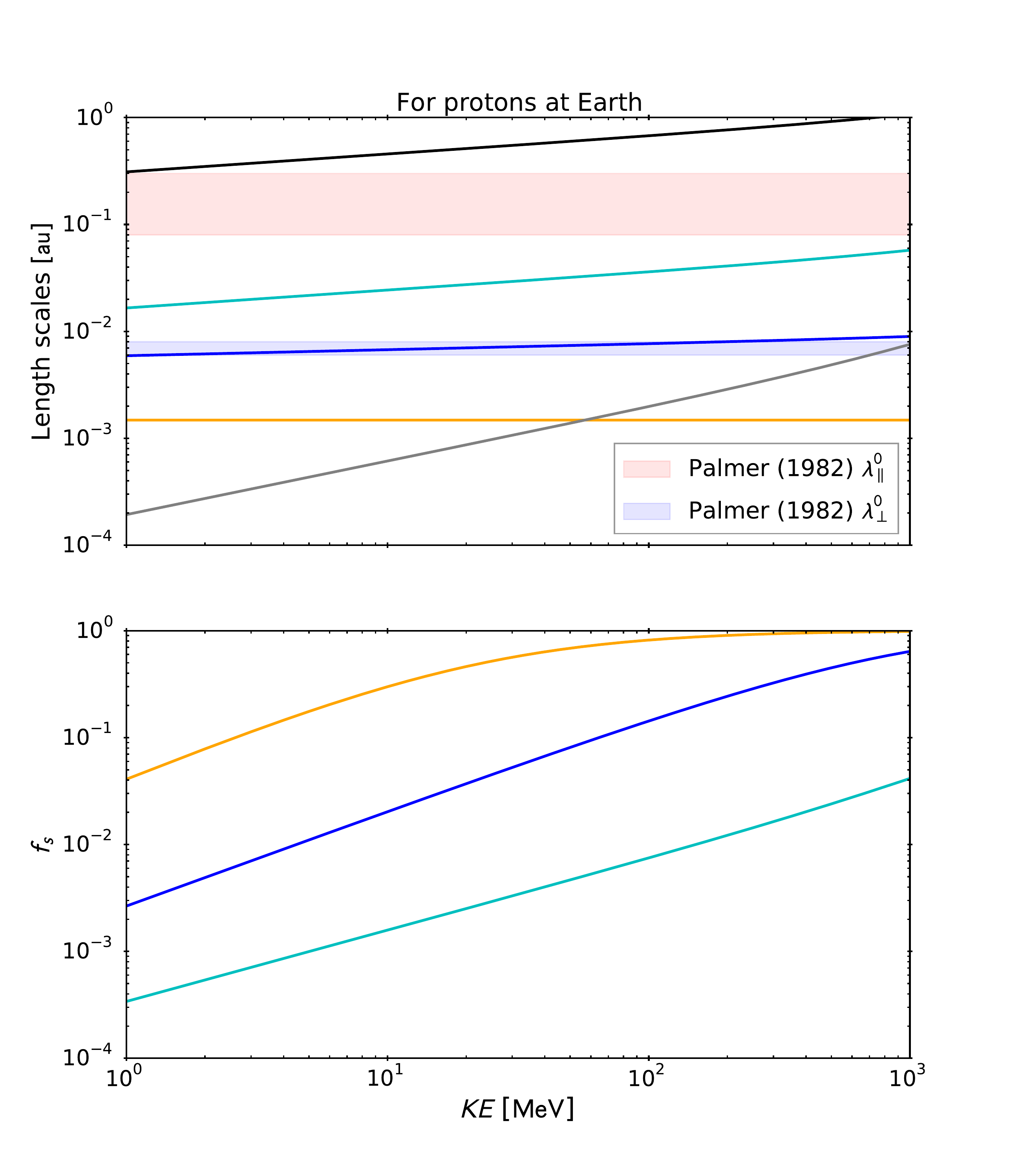}
\caption{\label{fig:Compare}\textit{Top panels:} Parallel and perpendicular mean free paths, together with the maximal Larmor radius. \textit{Bottom panels:} The resulting drift reduction factor (Eq.~\ref{eq:IsotropicDriftReduction}). \textit{Left panels:} As a function of radius for $100$ $\mathrm{MeV}$ protons using the turbulence transport model results in Fig.~\ref{fig:TTM}. \textit{Right panels:} As a function of kinetic energy for protons at Earth using observed turbulence quantities.}
\end{center}
\end{figure*}


\section{Consequences for Solar Energetic Particle Transport}
\label{sec:SEPs}

The influence of the behaviour of the various diffusion coefficients and the resulting drift reduction factors on SEP transport will be the subject of this section. Turbulence quantities and the perpendicular diffusion coefficient must be specified for Eq.~\ref{eq:MuDependentFs}. Appendix~\ref{apndx:Predict} discusses some key results of a turbulence transport model (TTM) in the very inner heliosphere, but simplified radial scalings will be used in this section for the turbulence quantities needed. The magnetic variance and correlation lengths of the slab and 2D turbulence are normalized to solar minimum observations at Earth, with the assumption of a slab/2D ratio of 20:80. Following \citet{burgeretal2008}, \citet{molotoetal2018}, or \citet{engelbrecht2019b}, the magnetic variance is modelled as $\delta B^2 = 10~{\rm nT}^2~(1~{\rm au}/r)^{2.4}$ \citep{bieberetal1994, zanketal1996, smithetal2001, smithetal2006, engelbrechtburger2013}, while the slab correlation length is modelled as $\ell_m = 2.5$~$\ell_{\rm 2D}$, with the 2D correlation length modelled by $\ell_{\rm 2D} = 7.41 \times 10^{-3}~{\rm au}~\sqrt{r / 1~{\rm au}}$ \citep{smithetal2001, weygandetal2011}. These scalings, illustrated in Fig.~\ref{fig:TTM}, are consistent with observations and TTMs in the heliosphere beyond Earth's orbit, but differ in the very inner heliosphere; the interested reader should consult Appendix~\ref{apndx:Predict} for these differences.

To investigate the turbulence-reduced drift coefficient, three different expressions derived using different theoretical approaches are employed for $D_{\perp}(\mu)$, which, according to Eq.~\ref{eq:perpMFPdef}, can be written as \cite[e.g.][]{engelbrecht2019b}
\begin{equation}
\label{eq:gmu}
D_{\perp} (\mu) = g(\mu) v \lambda_{\perp}^0,
\end{equation}
where $g(\mu)$ governs the perpendicular diffusion coefficient's pitch-angle dependence. In most of what is to follow, it will be assumed that $g(\mu)=\mu^2$, based on the numerical test particle simulations of \citet{casseetal2002} and the unified nonlinear theory (UNLT) of \citet{shalchi2010}. However, the consequences of assuming that $g(\mu)=|\mu|$, based on what is expected from the FLRW perpendicular diffusion coefficient \cite[see, e.g.,][]{jokipii1966, matthaeusetal1995, straussetal2017}, and an alternative form where $g(\mu)=\alpha \mu^2+\beta$ (with $\alpha$ and $\beta$ constants), motivated by the simulation results of \citet{qinshalchi2014} as discussed by \citet{engelbrecht2019b}, will also be considered. For the purposes of discussion, the choices of $D_{\perp}$ made below are motivated by their relative simplicity and tractability. More complicated expressions derived using, for example, more physically motivated forms for the 2D turbulence power spectra do exist in the literature \cite[see, e.g.,][]{engelbrechtburger2015, dempersengelbrecht2020}, but these are either nonlinear or depend on unusual functions (e.g. exponential integral and hypergeometric functions) that may complicate their characterization. Furthermore, such expressions often depend on physical quantities, such as the energy-range onset length scale \cite[see][]{matthaeusetal2007}, that are difficult to ascertain in the region of interest to this study \cite[see, e.g.,][]{engelbrecht2019a}.

\begin{figure*}
\begin{center}
\includegraphics[width=0.475\textwidth]{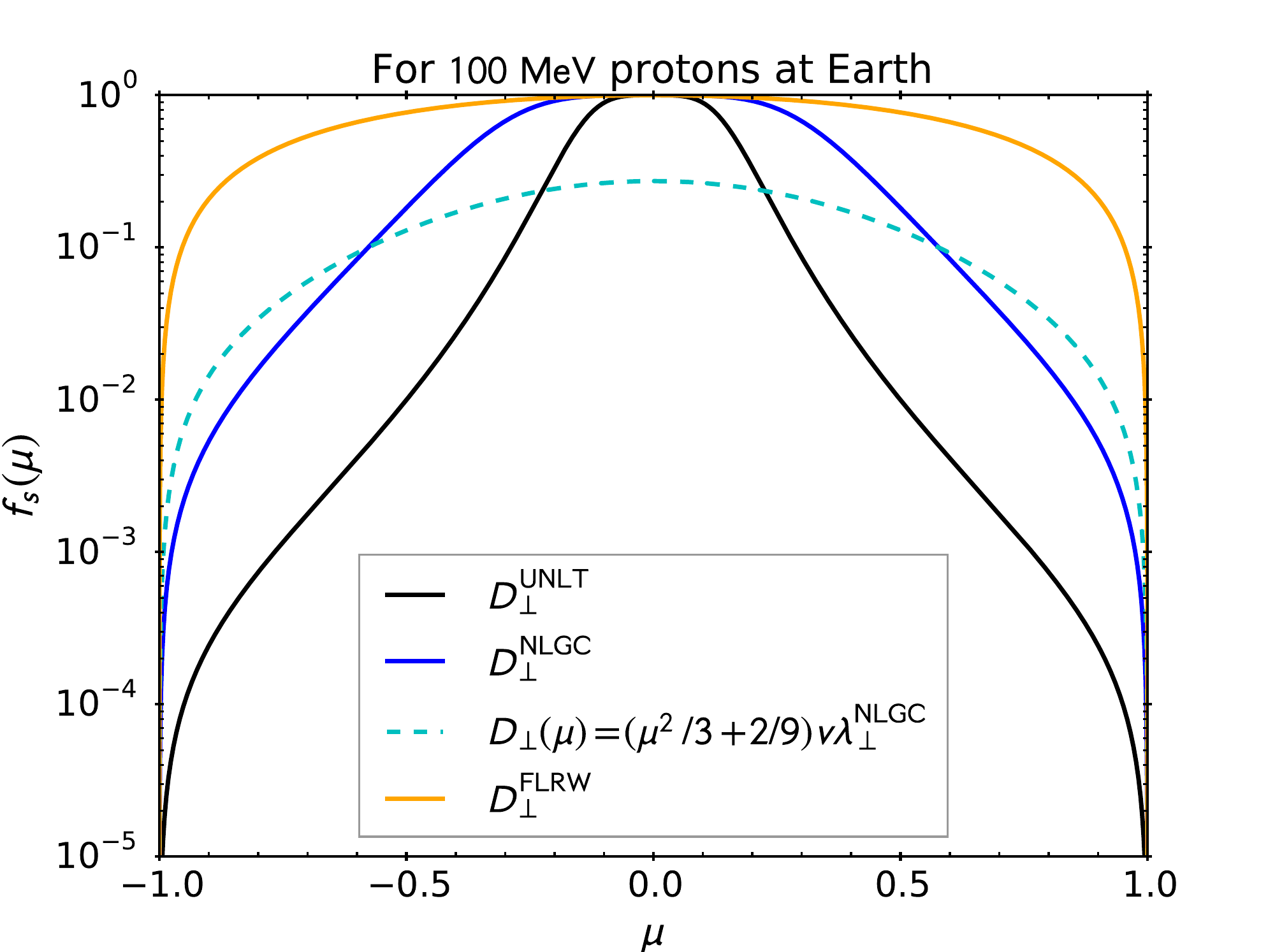}
\includegraphics[trim=40mm 15mm 25mm 15mm, clip, width=0.475\textwidth]{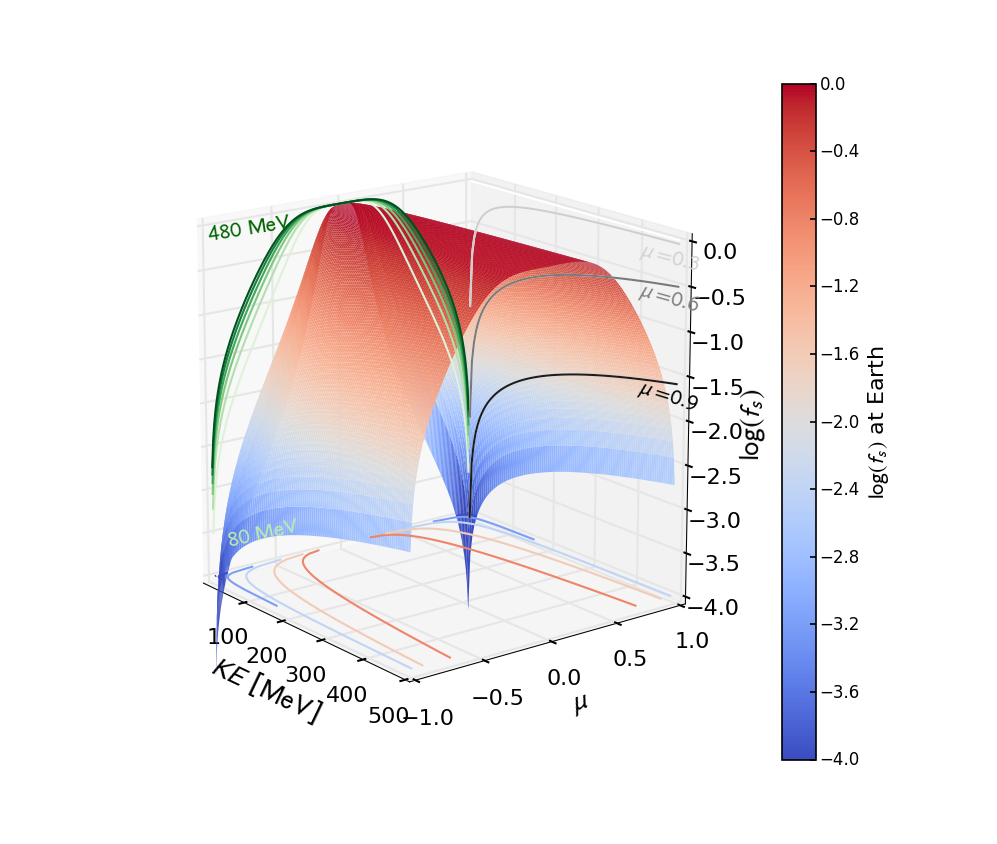}
\caption{\label{fig:EarthDependence}\textit{Left:} The pitch-angle dependent drift reduction factor (Eq.~\ref{eq:MuDependentFs}) for $100$ $\mathrm{MeV}$ protons at Earth predicted by different functional forms of the perpendicular diffusion coefficient. \textit{Right:} The pitch-angle dependent drift suppression factor (Eq.~\ref{eq:MuDependentFs}) for protons at Earth as a function of kinetic energy. Perpendicular diffusion is assumed to be described by the non-linear guiding centre theory (Eq.~\ref{eq:NLGC}).}
\end{center}
\end{figure*}

The first expression to be considered is based on an isotropic perpendicular MFP expression from the nonlinear guiding centre (NLGC) theory of \citet{matthaeusetal2003} derived by \citet{shalchietal2004} and modified for a general ratio of slab to 2D variance by \citet{burgeretal2008},
\begin{align}
\label{eq:NLGC}
D_{\perp} (\mu) & = \mu^2 v \lambda_{\parallel}^{1/3} \times \\
 & \;\;\;\; \left[ a^{2} \sqrt{3 \pi}\frac{2 \nu - 1}{\nu} \frac{\Gamma (\nu)}{\Gamma (\nu - 1/2)} \frac{\delta B_{\rm 2D}^2}{B_0^2} \ell_{2D} \right]^{2/3} \nonumber ,
\end{align}
where $\lambda_{\parallel}$ is the particle's parallel MFP and $a^2$ is assumed to be equal to $1/3$ following \citet{matthaeusetal2003}. The isotropic perpendicular MFP used in Eq.~\ref{eq:NLGC} is derived assuming a 2D turbulence power spectrum with a wave number-independent energy-containing range and an inertial range with spectral index $-2\nu$, which in this study is assumed to take on the Kolmogorov $-5/3$ value. Observationally, such a spectral form is not entirely realistic \cite[see, e.g.,][]{bieberetal1993, matthaeusetal2007}, but its use here is motivated by its simplicity relative to expressions derived using more realistic spectral forms \cite[see, e.g.,][]{engelbrechtburger2015, dempersengelbrecht2020}, as well as by the fact that its use in galactic cosmic ray modulation results has lead to computed intensities in good agreement with spacecraft observations \cite[e.g.][]{molotoengelbrecht2020, engelbrechtmoloto2021} and because it has also been applied to SEP modelling \citep[e.g.][]{qinetal2013, wangetal2014}.

The second expression for the pitch-angle dependent perpendicular diffusion coefficient to be used here is that derived by \citet{qinshalchi2014} from the \citet{shalchi2010} UNLT in the limit of dominant pitch-angle scattering. This expression is given by
\begin{equation}
\label{eq:UNLT}
D_{\perp} (\mu) = \frac{1}{2} a^2 \mu^2 v \frac{\delta B_{\rm 2D}^2}{B_0^2} \lambda_{\parallel}.
\end{equation}
The third pitch-angle dependent perpendicular diffusion coefficient to be considered here is the FLRW result, which can also be derived from UNLT \cite[e.g.][]{matthaeusetal1995, matthaeusetal2007, qinshalchi2014, straussetal2017}, and is given by
\begin{equation}
\label{eq:FLRW}
D_{\perp} (\mu) = a |\mu| v \kappa_{\rm FL},
\end{equation}
where $\kappa_{\rm FL}$ denotes the FLRW diffusion coefficient. The expression for this quantity employed here is that proposed by \citet{shalchiwienhorst2009}
\begin{equation}
\label{eq:KappaFL}
\kappa_{\rm FL} = \ell_{\rm 2D} \sqrt{\frac{2\nu - 1}{2 (q_{\rm 2D} - 1)} \frac{\delta B_{\rm 2D}^2}{B_0^2}} ,
\end{equation}
and is derived using the same 2D turbulence power spectrum form as used in Eq.~\ref{eq:UNLT}, with $q_{2D}>1$ denoting the energy-containing range spectral index.

Lastly, both the NLGC and UNLT expressions used here require an isotropic parallel MFP as input. To that end, an expression derived from quasi-linear theory (QLT) assuming magnetostatic slab turbulence by \citet{teufelschlickeiser2003} is employed, given by
\begin{equation}
\label{eq:LambdaParl}
\lambda_{\parallel} = \frac{3 s}{\pi (s - 1)} \frac{R^2}{k_m} \frac{B_0^2}{\delta B_{\rm sl}^2} \left[ \frac{1}{4} + \frac{2 R^{-s}}{(2 - s)(4 - s)} \right] ,
\end{equation}
where $R = R_L k_m$, with $k_m$ the wave number at which the inertial range, here assumed to have a spectral index of $-5/3=-s$, of the slab turbulence power spectrum commences.

The pitch-angle, energy, and radial dependencies of the turbulence-reduced drift reduction factor are discussed in detail in Appendix~\ref{apndx:Predict}, together with the radial and energy dependencies of the parallel and different perpendicular MFPs discussed above. The bottom two panels of Fig.~\ref{fig:Compare} show Eq.~\ref{eq:IsotropicDriftReduction} for protons, calculated using the MFPs in the top two panels, as function of radial distance (left panel) and kinetic energy (right panel). Larger values for $\lambda_{\perp}^0$ lead to smaller values for $f_s$ and hence, a greater reduction in drift effects. As function of radial distance, the $100$~${\rm MeV}$ proton drift reduction function is relatively flat, only showing a moderate upturn at the smallest radial distances considered here for the UNLT perpendicular MFP. Almost no drift reduction can be seen at this energy for FLRW, but this is not the case for all energies, as $f_s$ calculated using the FLRW MFP does begin to drop below $1$ at energies below $\sim 100$~${\rm MeV}$. The UNLT result remains well below unity over the entire radial and energy range considered here.

The left panel of Fig.~\ref{fig:EarthDependence} shows the pitch-angle dependence of Eq.~\ref{eq:MuDependentFs} for $100$~${\rm MeV}$ protons and typical turbulence quantities measured at Earth. Field aligned particles ($\mu = \pm 1$) will have a significant reduction in their drift velocities for all of the different forms of $D_{\perp} (\mu)$ considered, while the behaviour of $f_s(\mu)$ towards $\mu = 0$ depends critically on the theory used for $D_{\perp} (\mu)$. Any perpendicular diffusion coefficient which is zero at $\mu = 0$ will have $f_s(0) = 1$, and hence particles will experience full drift effects at this pitch-angle, but there does not seem to be any \textit{a priori} reason that a particle with a $90^{\circ}$ pitch-angle should not experience drift reductions. Both the pitch-angle and energy dependence of Eq.~\ref{eq:MuDependentFs} are shown in the right panel of Fig.~\ref{fig:EarthDependence} at Earth for NLGC. For a fixed pitch-angle (except at $\mu = 0$), $f_s$ becomes more pronounced at low energies. The main behaviour of $f_s(\mu)$ is as follows: field aligned particles will experience more drift reductions, low energy particles will experience less drifts, and drift reduction factors will have a weak radial dependence.

\begin{table*}
\caption{\label{tab:SimLabels} Simulation naming}
\centering                                      
\begin{tabular}{c l l l}  
\hline\hline                        
Simulation number & Simulation label & Transport processes included in the simulation\\    
\hline                                   
    1 & $\parallel$ & only parallel transport \\      
    2 & $\parallel, V_d$ & parallel transport and (non-reduced) drifts\\
    3 & $\parallel, V_d$, FLRW & parallel transport, drifts, and FLRW (see Eq.~\eqref{eq:FLRW})  \\
    4 & $\parallel, f_sV_d$, FLRW & parallel transport, reduced drifts, and FLRW \\
    5 & $\sqrt{f_s}\parallel, f_sV_d$, FLRW & reduced parallel transport, reduced drifts, and FLRW \\
    6 & $\parallel, V_d$, NLGC & idem to 3 but for NLGC (see Eq.~\eqref{eq:NLGC}) \\
    7 & $\parallel, f_sV_d$, NLGC & idem to 4 but for NLGC     \\
    8 & $\sqrt{f_s}\parallel, f_sV_d$, NLGCT & idem to 5 but for NLGC  \\
    9 & $\parallel, V_d$, UNLT & idem to 3 but for UNLT (see Eq.~\eqref{eq:UNLT}) \\
    10 & $\parallel, f_sV_d$, UNLT &  idem to 4 but for UNLT   \\
    11 & $\sqrt{f_s}\parallel, f_sV_d$, UNLT &idem to 5 but for UNLT \\
\hline                                             
\end{tabular}
\end{table*}

To illustrate the possible effect of the reduction factor $f_s(\mu)$ on SEP events, Eq.~\ref{eq:MuDependentFs} is incorporated into the energetic particle model PARADISE \citep[][]{wijsenetal2019,wijsen2020}. 
This model solves the focused transport equation, including the effect of cross-field diffusion and particle drifts. 
In this work, protons are simulated propagating trough a Parker SW of constant radial speed $v_{\rm sw} = 400$~${\rm km} \cdot {\rm s}^{-1}$. Moreover, the HMF is assumed to have a positive polarity ($A = +1$) and is scaled such that $B(1~{\rm \, au}) = 5$~${\rm nT}$. The positive HMF polarity implies that the protons will drift in the southward direction \citep[see e.g.][]{dallaetal2013}. The functional form of the pitch-angle diffusion coefficient is derived from QLT and scaled such that the isotropic parallel MFP corresponds to Eq.~\eqref{eq:LambdaParl}. For the simulations including cross-field diffusion, the three different functional forms introduced in Eq.~\ref{eq:NLGC}, Eq.~\ref{eq:FLRW}, and Eq.~\ref{eq:UNLT} are considered.

As a boundary condition for the energetic particles, the isotropic injection function
\begin{align*}
f(r,\theta,\phi,E,\mu,t) \propto & \frac{1}{t} \exp{\left( - \frac{\tau_a}{t} - \frac{t}{\tau_e} \right)} \exp{\left[ \frac{\phi^2 + \theta^2}{2 \sigma^2} \right]}\\
 & \times \delta(E/E_0 - 1) \delta(r/r_0 - 1)
\end{align*}
is prescribed at $r_0 = 0.05$~${\rm au}$. A Gaussian injection profile with $\sigma = 5^\circ$ is assumed for the azimuthal $\phi$ and the latitudinal $\theta$ angles, with the injection centred on the ecliptic plane. The injection function follows a Reid–Axford temporal profile \citep{reid1964}, with $\tau_a = 0.2$~${\rm h}$ and $\tau_e = 1$~${\rm h}$. The kinetic energy $E$ of the injected protons is assumed to be $E_0 = 100$~${\rm MeV}$. The particles are propagated for a total simulation time of $T= 50$~${\rm h}$.

An overview of the different simulations that were performed can be found in Table~\ref{tab:SimLabels}. In what follows the different simulations will be referred to according to their number in Table~\ref{tab:SimLabels}.
For each of the simulations, the energy-integrated omni-directional intensity,
\begin{equation*}
\label{eq:ODI}
I(r,\theta,\phi,t) = \int_0^{E_0} \int_{-1}^1 j(r,\theta,\phi,E,\mu,y) d\mu \, dE
\end{equation*}
with $j = p^2 f$ denoting the differential intensity, were calculated, together with the ratio
\begin{equation}
\label{eq:Ratio}
R(r,\theta,\phi) := \frac{\int_0^T I(r,\theta,\phi,t) dt}{\int_0^T I(r,-\vartheta,\phi,t) dt} .
\end{equation}

\begin{figure}
\centering
\includegraphics[width=0.47\textwidth]{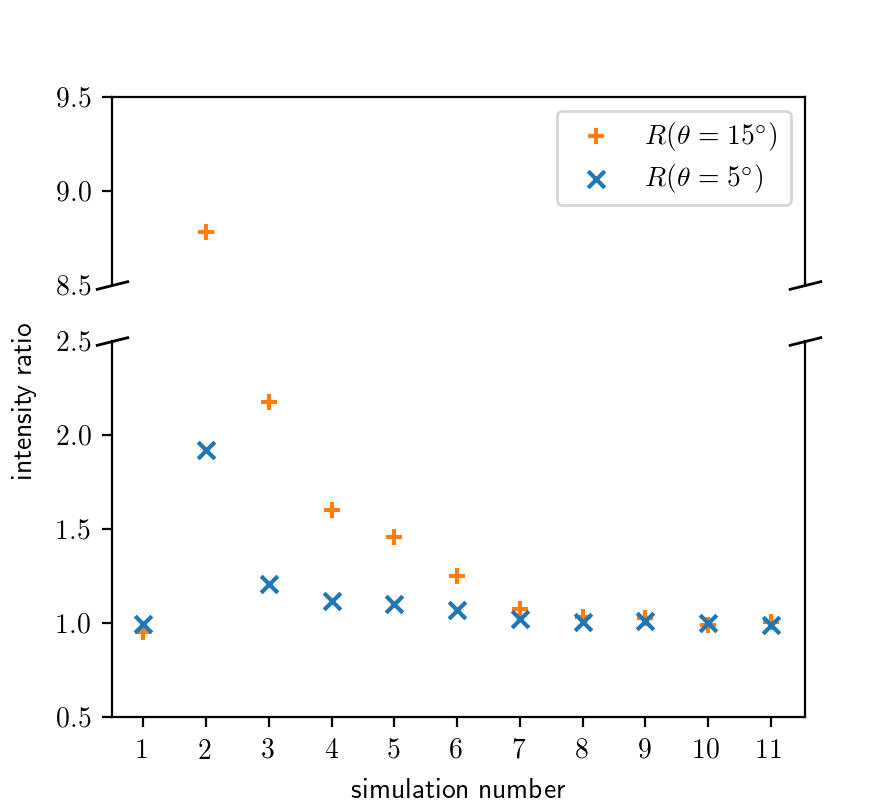}
\caption{Event-integrated intensity ratios $R(\theta = 15^\circ)$ and $R(\theta = 5^\circ)$ (Eq.~\ref{eq:Ratio}) for observers located at $r= 1$~${\rm au}$ and magnetically connected to $\theta = 0^\circ$ and $\phi = 0^\circ$ at the inner boundary. Descriptions of the simulation numbers on the x-axis can be found in Table~\ref{tab:SimLabels}.}
\label{fig:Ratios}
\end{figure}

In particular, the ratios $R(\theta = 15^\circ)$ and $R(\theta = 5^\circ)$ for observers located at $r = 1$~${\rm au}$ and magnetically connected to $\theta = 0^\circ$ and $\phi = 0^\circ$ at the inner boundary are considered. Fig.~\ref{fig:Ratios} shows the values obtained for these two ratios for the different simulation. Without drifts, the particle distribution will remain symmetric  with respect to the injection site and hence $R(\theta = 15^\circ) = R(\theta = 5^\circ) = 1$. When drifts are included in the simulations, protons drift southwards and as a consequence, $R(\theta = 15^\circ) > R(\theta = 5^\circ) > 1$. Fig.~\ref{fig:Ratios} illustrates that when no perpendicular diffusion or drift reduction is present, the effect of drifts can be significant, since $R(\theta = 15^\circ) = 8.8$ and  $R(\theta = 5^\circ) = 1.9$ for simulation~2. Note that the drift effects are only seen very late in the event ($> 2$~${\rm h}$) and that these ratios use the event integrated omni-directional intensity to clearly illustrate differences. Fig.~\ref{fig:Ratios} also illustrates that these ratios are drastically reduced when cross-field diffusion is switched on (simulations~3, 6 or 9). This reduction is the largest for the simulations with UNLT and the smallest for the simulations with FLRW. This ordering follows the ordering of the associated perpendicular MFPs, i.e., $\lambda_\perp^{\rm UNLT} > \lambda_\perp^{\rm NLGC} > \lambda_\perp^{\rm FLRW}$ (see Fig.~\ref{fig:Compare}). In fact, under the assumed turbulence conditions, the UNLT cross-field diffusion is strong enough to erase the drift-induced asymmetries almost entirely. When the drift reduction factor $f_s$ is taken into account (simulations~4 or 7), the remaining drift-induced asymmetries in the simulations with NLGC theory and FLRW decrease further, as expected. Moreover,  Fig.~\ref{fig:Ratios} illustrates that the ratios decrease even further when also taking into account the reduction of the particle streaming along the HMF (e.g., simulations~5 or 8). This is because this reduced parallel streaming keeps the particles for a prolonged amount of time at smaller heliocentric radial distances where the drifts are weaker \citep[e.g.,][]{dallaetal2013}.

\begin{figure}
\centering
\includegraphics[width=0.45\textwidth]{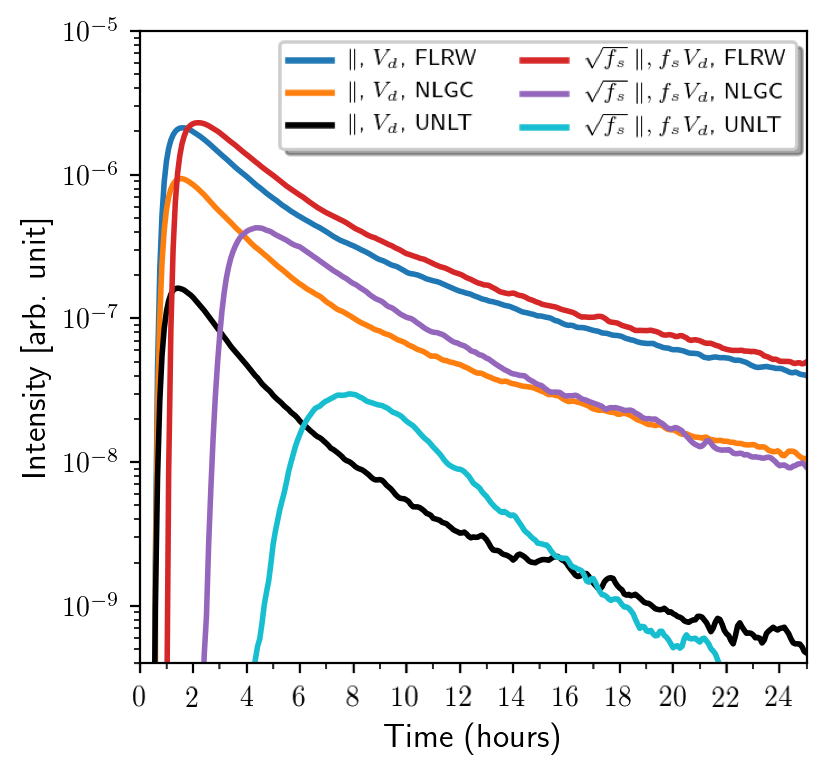}
\caption{Omni-directional intensity profiles as a function of time for an observer located at $r= 1$~${\rm au}$ and magnetically connected to the center of the injection region.}
\label{fig:Intensities}
\end{figure}

\begin{figure*}
\centering
\includegraphics[width=0.85\textwidth]{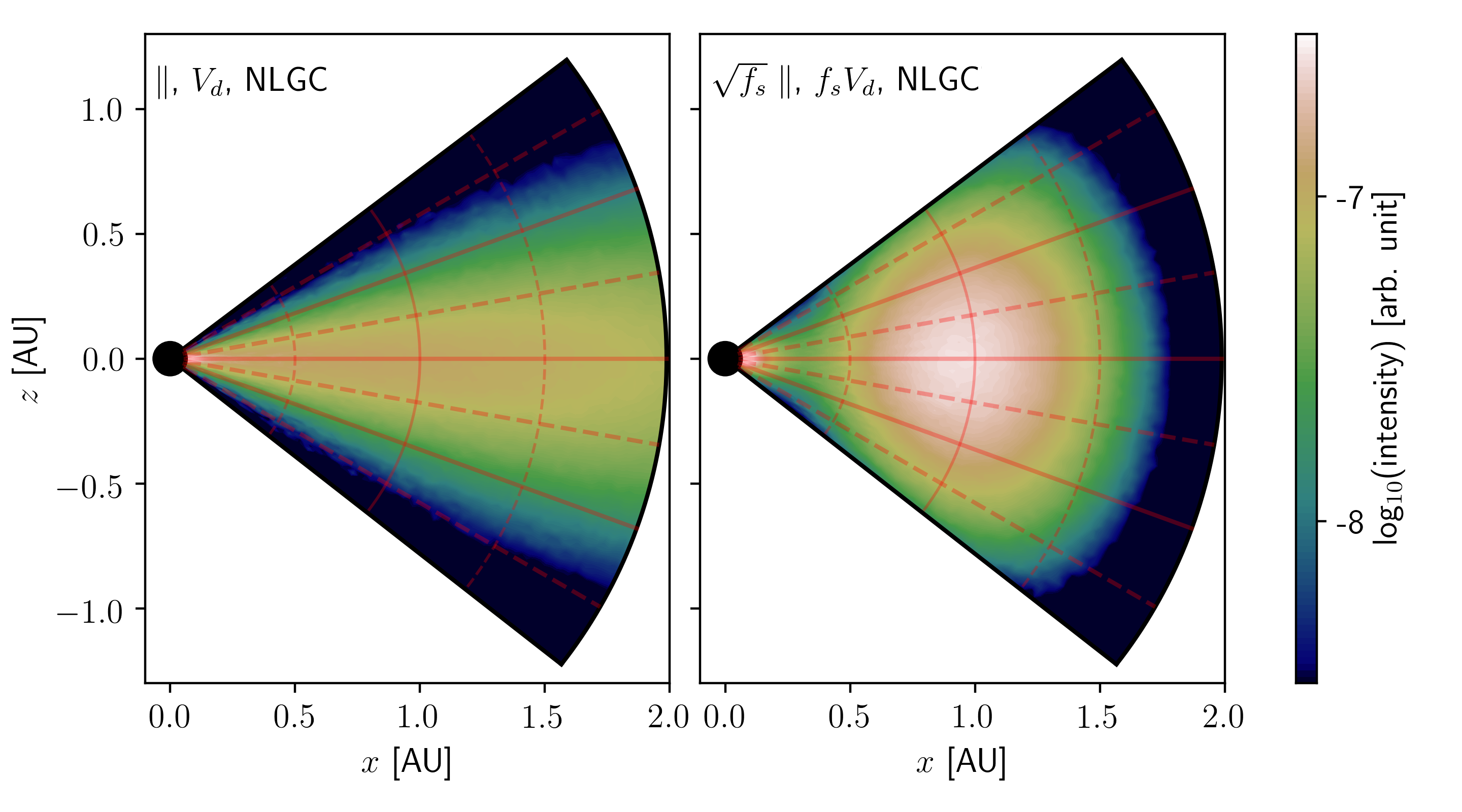}
\caption{Omni-directional intensity at time $t= 5$~${\rm h}$, measured along the magnetic field lines with foot points located at $\phi = 0^{\circ}$ and projected on a meridional slice. \textit{Left panel:} Without a reduction in the streaming term. \textit{Right panel:} If the reduction of the streaming term is included.}
\label{fig:Meridionals}
\end{figure*}

Fig.~\ref{fig:Intensities} shows the omni-directional time-intensity profiles for simulations 3, 5, 6, 8, 9 and 11 for an observer located at $r = 1$~${\rm au}$ and magnetically connected to the center of the injection region. It can be seen that the reduction factor $\sqrt{f_s}$ on the streaming of the protons along the HMF delays the onset times by $\sim 0.5$~${\rm h}$, $\sim 2.8$~${\rm h}$ and $\sim 3.5$~${\rm h}$ for FLRW, NLGC, and UNLT, respectively. This is in accordance with $\sqrt{f_s(\mu)}$ being smallest for UNLT and largest for FLRW (see Figs.~\ref{fig:Compare} and~\ref{fig:EarthDependence}). 
Since $f_s(\mu)$ becomes smaller when $|\mu|$ approaches unity, the spreading of the particles along the field lines will be reduced. As a result, the injected distribution function will propagate as a more coherent pulse along the HMF. This is illustrated in Fig.~\ref{fig:Meridionals}, which shows for simulations 6 and 7 the omni-directional intensity at time $t = 5$~${\rm h}$ measured along the magnetic field lines with foot points at $\phi = 0^{\circ}$ and projected on a meridional slice.

\section{Summary and Discussion}
\label{sec:Discuss}

This paper investigated the possible pitch-angle dependence of the turbulent drift reduction factor and its implications for SEP drifts. Section~\ref{sec:Intro} introduced the current application of drifts to SEPs and the major shortcomings of these efforts by not considering drift reductions. The investigation began in Section~\ref{sec:Reduce} with a summary of why drift reduction occurs and what is known about drift suppression in the isotropic case, which was supplemented by a conceptual derivation of the drift reduction in Appendix~\ref{apndx:Derivation}. Here it was emphasized that there is an interplay between drifts and turbulence: in the absence of turbulence, particles experience drift motions, but turbulence disrupts these drift patterns and cause the particles to rather diffuse perpendicular to the background magnetic field.

A pitch-angle dependent drift reduction factor was also derived in Section~\ref{sec:Reduce}. The derived form is able to describe full orbit simulations qualitatively, but is critically dependent on the assumed pitch-angle dependence of the perpendicular diffusion coefficient. A perpendicular diffusion coefficient which is zero for particles with $90^{\circ}$ pitch-angles will yield no drift reductions for these particles, while perpendicular diffusion coefficients which is a maximum for field aligned particles will yield no drifts for these particles. Theories of cross-field diffusion predicting larger perpendicular MFPs will also yield more drift reductions \citep[in agreement with][]{engelbrechtetal2017}. The drift reduction factor was then evaluated for realistic turbulence and scattering conditions in Appendix~\ref{apndx:Predict} to predict its effects on SEP propagation. Here it was found that field aligned particles will experience more drift reductions, that low energy particles will experience less drifts, and that drift reduction factors will have a weak radial dependence.

The drift reduction was investigated with the PARADISE model \citep{wijsen2020} for protons subject to perpendicular diffusion coefficients with different pitch-angle dependencies originating from the different FLRW, NLGC, and UNLT theoretical approaches (see the discussion in Section~\ref{sec:SEPs}). Although drifts could potentially have a large effect, it was seen that the inclusion of cross-field diffusion in a realistic scenario already diminishes the effect of drifts (see Fig.~\ref{fig:Ratios}). In the simulations, the strong perpendicular diffusion predicted by UNLT, almost smeared out the signatures of drifts completely, while the weaker perpendicular diffusion predicted by employing the smaller FLRW perpendicular diffusion coefficient still showed signatures of drifts. By including the proposed turbulence-reduced pitch-angle dependent drift coefficient into the simulations, drift reduction then further diminishes the effect of drifts, such that drifts can be considered to be a second order effect. It was also seen that drifts are further reduced when the streaming of particles, which will be discussed further below, is also included. This is because particles then spend more time closer to the Sun where drift effects are smaller. For realistic turbulence conditions in the heliosphere, as considered here, it can therefore be expected that cross-field diffusion will have the largest influence on the perpendicular transport of SEPs and not particle drifts. Note that a Parker HMF was used in these modelling efforts and that drifts in the geometry of magnetic field structures, like coronal mass ejections or co-rotating interaction regions, might still be important \citep[according to the modelling of][]{wijsenetal2020}, although it would also be dependent on the turbulence conditions and levels, which could be significantly enhanced in these structures \cite[see, e.g.,][]{wiengartenetal2015}.

It is interesting to note the prediction in Eq.~\ref{eq:ParkerDriftVelocity} that the parallel streaming term ($\mu v \, \hat{b}_0$) will also be reduced. Modelling showed that this reduction in streaming will cause a delayed onset at an observer (Fig.~\ref{fig:Intensities}), with the delay larger for longer perpendicular MFPs. When combined with perpendicular diffusion, this might solve the causality problem identified by \citet{straussfichtner2015}, because particles with $\mu = \pm 1$ will stream at a speed less than $v$, allowing them to also diffuse (if $D_{\perp} (\mu = \pm 1) \neq 0$) without moving faster than $v$ (depending on the value of $D_{\perp}$). This factor is also comparable to the factor (the cosine of the angle between the local mean magnetic field and the local random walking magnetic field) used by \citet{laitinenetal2013} to have particles following their computed random walking field lines. Hence, this will have an influence on the analysis of \citet{laitinendalla2019}, where the path length of a diffusing particle is calculated (with the causality problem also present in some approaches), and it would explain why these authors argue that the streaming should be re-scaled. Lastly, this may also influence SEP path lengths inferred from, e.g., PSP/IS{\ensuremath{\odot}}IS observations, as is done by \citet{chhiberetal2021}.

A type of drift which has not yet been addressed in this work is that of particle drifts along a current sheet. \citet{burger2012} derived an expression for the neutral sheet drift velocity in terms of the gradient and curvature drift for a nearly isotropic distribution \citep[see also][]{engelbrechtetal2019}. Although this approach does not contain a pitch-angle dependence, it yields a result dependent on the drift coefficient and one could therefore expect that neutral sheet drifts will be reduced by the same factor as the gradient and curvature drifts. The findings of \citet{battarbeeetal2017}, that particles have difficulty crossing a current sheet, seem to be verifiable with observations \citep[the observations of][might point to this]{klassenetal2018}. Notice, however, that such an observation is still possible even if current sheet drifts are suppressed, because the current sheet could act as a diffusion barrier, as discussed by \citet{straussetal2016}, due to, e.g., lower turbulence conditions close to the heliospheric current sheet \citep[as observed by PSP;][]{chenetal2021}. Furthermore, given the observed complexity of current sheets and the plasma conditions in their vicinity \cite[for a review, see][]{khabarovaetal2021}, modelling the transport of charged particles in their vicinity presents unique challenges \cite[see, e.g.,][]{pezzietal2021} that are beyond the scope of this work.

The new pitch-angle dependent drift reduction factor proposed here, as opposed to the isotropic result of \citet{engelbrechtetal2017}, cannot as yet be compared with the results of existing numerical test particle simulations of the particle drift coefficients such as those performed by, e.g., \citet{tautzshalchi2012}, as no direct simulations investigating the pitch-angle dependence of this drift reduction factor currently, to the best of our knowledge, exist. Therefore, although the derived pitch-angle dependent drift reduction factor, when averaged over pitch-angle, coincides nicely with the isotropic drift reduction factor derived by \citet{engelbrechtetal2017} for weak turbulence conditions and small perpendicular MFPs, it is unclear whether this reduction factor's form would generally hold for all turbulence conditions expected of the very inner heliosphere. This is evinced by the fact that for large perpendicular MFPs it was found that the pitch-angle dependent drift reduction factor is larger than the isotropic factor, depending on the functional form used for the perpendicular diffusion coefficient (see Fig.~\ref{fig:ValidityRange}). It is possible that this functional form, derived as it is from the \citet{biebermatthaeus1997} theory with its implicit assumption of low turbulence levels, would simply not be applicable under the higher turbulence levels which would theoretically \cite[e.g.][]{shalchi2009} lead to larger perpendicular MFPs. Furthermore, the consequences of the pitch-angle dependence derived here for the drift reduction factor, such as the prediction that particles with $90^{\circ}$ pitch-angles will experience no drift reductions if the perpendicular diffusion coefficient is zero for that pitch-angle \cite[e.g.][]{engelbrecht2019b}, would need to be investigated by means of direct numerical test particle simulations. This will be the subject of future work.


\acknowledgments

This work is based on the research supported in part by the National Research Foundation of South Africa (NRF grant numbers 120847, 120345, and 119424). Opinions expressed and conclusions arrived at are those of the authors and are not necessarily to be attributed to the NRF. JPvdB acknowledge support from the South African National Space Agency. NW acknowledges funding from the Research Foundation - Flanders (FWO -- Vlaanderen, fellowship no. 1184319N). Computational resources and services used in this work were provided by the VSC (Flemish Supercomputer Centre), funded by the FWO and the Flemish Government-Department EWI. 
Figures prepared with Matplotlib \citep{hunter2007} and certain calculations done with NumPy \citep{harrisetal2020} and SciPy \citep{virtanenetal2020}.


\appendix

\section{Drift Reduction in the Isotropic and Anisotropic Limit}
\label{apndx:Derivation}

In galactic cosmic ray modulation studies it is customary to absorb the pitch-angle averaged GC drift velocity,
\begin{equation}
\label{eq:IsotropicDriftVel}
\left\langle \vec{v}_d \right\rangle = \frac{v p}{3 q} \vec{\nabla} \times \left( \frac{\hat{b}_0}{B_0} \right) ,
\end{equation}
into the diffusion tensor as a drift coefficient, i.e. $\left\langle \vec{v}_d \right\rangle = \vec{\nabla} \times (\kappa_A \hat{b}_0)$, with the subscript A referring to the fact that it forms the antisymmetric part of the diffusion tensor \cite[see e.g.,][]{burgeretal2008}. This can be done if the GC velocity is divergence free, which Eq.~\ref{eq:IsotropicDriftVel} is in the absence of electric fields \citep{minnieetal2007, burgervisser2010, engelbrechtetal2017}. An example of how the drift coefficient is calculated following \citet{biebermatthaeus1997} and \citet{burgervisser2010} will be briefly discussed. The drift coefficient is calculated from a velocity correlation according to the Taylor-Green-Kubo (TGK) formalism \cite[see][and references therein]{shalchi2009},
\begin{equation}
\label{eq:DxyDefine}
D_{xy} (\mu) = \int_0^{\infty} \langle v_x(0) v_y(t) \rangle dt ,
\end{equation}
where $D_{xy} (\mu)$ is further averaged over pitch-angle to give $\kappa_A = \int_{-1}^1 D_{xy} (\mu) d\mu / 2$.

Consider a homogeneous background magnetic field along the $z$-axis. It can be showed that the particle's velocity components perpendicular to the magnetic field can be written as \citep{biebermatthaeus1997}
\begin{align*}
v_x(t) & = v_{\perp} \cos (\omega_c t - \varphi) \\
v_y(t) & = - {\rm sign}(q) v_{\perp} \sin (\omega_c t - \varphi) ,
\end{align*}
where $\varphi$ is the particle's gyro-phase and $v_{\perp} = v \sqrt{1 - \mu^2}$ is the particle's speed perpendicular to the magnetic field. Now it can be calculated that
\begin{equation*}
v_x(0) v_y(t) = {\rm sign}(q) v_{\perp}^2 \left[ \cos (\omega_c t) \sin \varphi \cos \varphi - \sin (\omega_c t) \cos^2 \varphi \right]
\end{equation*}
and this can be averaged over gyro-phase to obtain
\begin{equation*}
\langle v_x(0) v_y(t) \rangle_{\varphi} = - {\rm sign}(q) \frac{1 - \mu^2}{2} v^2 \sin (\omega_c t) = - \langle v_y(0) v_x(t) \rangle_{\varphi} ,
\end{equation*}
which also shows why the drift coefficient forms the antisymmetric part of the diffusion tensor \citep{burgervisser2010}.

\citet{biebermatthaeus1997} argued that turbulence (assumed to be stationary and homogeneous) will decorrelate the unperturbed velocity components, such that the correlation tends to zero in a finite time (i.e. as time $t \longrightarrow \infty$, the velocity correlation functions approach zero faster than $1/t$), and assumed that the correlation decreases exponentially with a characteristic time $\tau$, such that the particle distribution becomes nearly isotropic. The drift coefficient can then be calculated from Eq.~\ref{eq:DxyDefine} as
\begin{equation*}
D_{xy} (\mu) \approx - {\rm sign}(q) \frac{1 - \mu^2}{2} v^2 \int_0^{\infty} \sin (\omega_c t) e^{- t / \tau} dt = - {\rm sign}(q) \frac{1 - \mu^2}{2} v^2 \frac{\omega_c}{1 / \tau^2 + \omega_c^2} = - \frac{1 - \mu^2}{2} 3 \kappa_A^{\rm ws} f_s ,
\end{equation*}
where
\begin{equation}
\label{eq:WeakScatteringLimit}
\kappa_A^{\rm ws} = \frac{v p}{3 q B_0} = {\rm sign}(q) \frac{v R_L}{3}
\end{equation}
is the weak scattering drift coefficient \cite[see][]{formanetal1974} and 
\begin{equation*}
f_s = \frac{(\omega_c \tau)^2}{1 + (\omega_c \tau)^2} .
\end{equation*}
is the so-called drift reduction factor. Since an unperturbed orbit is assumed, the current derivation is strictly only valid for low turbulence levels. Notice that $\kappa_A$ will be non-zero in a homogeneous field without turbulence ($\tau \longrightarrow \infty$), due to the velocity correlations in Eq.~\ref{eq:DxyDefine}, but that $\left\langle \vec{v}_d \right\rangle$ will then be zero because $\kappa_A^{\rm ws} \hat{b}_0$ will be constant \cite[see][]{burgervisser2010}.

\citet{engelbrechtetal2017} present a conceptual explanation of the origins of drift suppression which will be generalised here \cite[see also][]{zhaoetal2017}. Consider the gyro-phase averaged GC velocity in a flow frame \citep[like the SW; see][for the derivation]{zhang2006, wijsen2020}
\begin{align}
\label{eq:GyroAvgGCVel}
\langle \vec{v}_{\rm gc} \rangle_{\varphi} & = \mu v \, \hat{b} + \vec{w} + \frac{m}{qB} \hat{b} \times \left[ \frac{\partial \vec{w}}{\partial t} + (\mu v \, \hat{b} + \vec{w}) \cdot \vec{\nabla} \vec{w} \right] + \frac{\mu p}{qB} \hat{b} \times \left( \frac{\partial \hat{b}}{\partial t} + \vec{w} \cdot \vec{\nabla} \hat{b} \right) + \nonumber \\
 & \;\;\;\; \frac{vp}{qB} \left\lbrace \mu^2 (\vec{\nabla} \times \hat{b})_{\perp} + \frac{1 - \mu^2}{2} \left[ (\vec{\nabla} \times \hat{b})_{\parallel} + \frac{\hat{b} \times \vec{\nabla} B}{B} \right] \right\rbrace + \nonumber \\
 & \;\;\;\; \frac{vp}{c^2qB} \left\lbrace \mu^2 \left( \vec{w} \times \frac{\partial \hat{b}}{\partial t} \right)_{\perp} +  \frac{1 - \mu^2}{2} \left[ \left( \vec{w} \times \frac{\partial \hat{b}}{\partial t} \right)_{\parallel} + \frac{\partial B}{\partial t} \frac{\hat{b} \times \vec{w}}{B} \right] \right\rbrace
\end{align}
where $\vec{w}$ is the flow velocity, the particle's momentum, speed, and pitch-angle are all measured in the flow frame, the fields are evaluated at the particle's position in the observer's frame \citep[note that if the fields are evaluated at the GC's position, then the $(\vec{\nabla} \times \hat{b})_{\parallel}$ and $(\vec{w} \times \partial \hat{b} / \partial t)_{\parallel}$ terms will be absent;][]{rossiolbert1970, burgeretal1985, wijsen2020}, $m$ is the particle's relativistic mass in the flow frame, and the subscripts $\perp$ and $\parallel$ indicate directions perpendicular and parallel to $\hat{b}$, respectively. It is important to notice that this equation is derived assuming that the fields do not vary much over a Larmor radius, that is $B / |\partial B_i / \partial x_j| \gg R_L$ and $w / |\partial w_i / \partial x_j| \gg R_L$ \citep{rossiolbert1970, burgeretal1985, wijsen2020}.

This equation is normally defined with respect to a smooth background magnetic field, but consider what would happen if it were defined with respect to the total (background and turbulent) magnetic field. This approach might raise some problems: 1) the instantaneous GC position might not be well defined in a turbulent magnetic field, especially if the turbulence is strong, and this instantaneous GC velocity might not be valid \citep[see the discussions of][]{burgeretal1985}; 2) the pitch-angle might also be ill defined if it is with respect to the total magnetic field. Nonetheless, one can hope that if suitable averages are taken over the magnetic field and if the turbulence is not too strong, it would yield well behaved quantities. For weak turbulence conditions, it will then be explicitly assumed that the pitch-angle is defined with respect to the background magnetic field.

Assume that the fields can be written as the sum of a background and fluctuating component, i.e. $\vec{B} = \vec{B}_0 + \delta \vec{B}$, where $\langle \vec{B} \rangle = \vec{B}_0$ for a suitable average, and that the fluctuations are transverse, such that $\vec{B}_0 \cdot \delta \vec{B} = 0$. The total magnetic field magnitude is then $B = B_0 \sqrt{1 + \delta B^2 / B_0^2}$, where $\delta B^2 \ll B_0^2$ for weak turbulence, and with velocity fluctuations, the flow velocity would be $\vec{w} = \vec{w}_0 + \delta \vec{w}$. Substituting this into the GC velocity, allows it to be decomposed into an average and perturbed motion, i.e. $\langle \vec{v}_{\rm gc} \rangle_{\varphi} = \vec{v}_{\rm gc} + \delta \vec{v}_{\rm gc}$,
\begin{align*}
\vec{v}_{\rm gc} & = \xi \mu v \, \hat{b}_0 + \vec{w}_0 + \frac{\xi^2 m}{q B_0} \, \hat{b}_0 \times \left[ \frac{\partial \vec{w}_0}{\partial t} + (\xi \mu v \, \hat{b}_0 + \vec{w}_0) \cdot \vec{\nabla} \vec{w}_0 \right] + \frac{\xi^3 \mu p}{q B_0} \, \hat{b}_0 \times \left( \frac{\partial \hat{b}_0}{\partial t} + \vec{w}_0 \cdot \vec{\nabla} \hat{b}_0 \right) + \\
 & \;\;\;\;\; \frac{\xi^2 v p}{q B_0} \left\lbrace \mu^2 (\vec{\nabla} \times \hat{b}_0)_{\perp} + \frac{1 - \mu^2}{2} \left[ (\vec{\nabla} \times \hat{b}_0)_{\parallel} + \frac{\hat{b}_0 \times \vec{\nabla} B_0}{B_0} \right] \right\rbrace + \\
 & \;\;\;\;\; \frac{\xi^2 v p}{c^2 q B_0} \left\lbrace \mu^2 \left( \vec{w}_0 \times \frac{\partial \hat{b}_0}{\partial t} \right)_{\perp} + \frac{1 - \mu^2}{2} \left[ \left( \vec{w}_0 \times \frac{\partial \hat{b}_0}{\partial t} \right)_{\parallel} +  \frac{\partial B_0}{\partial t} \frac{\hat{b}_0 \times \vec{w}_0}{B_0} \right] \right\rbrace
\end{align*}
\begin{align*}
\delta \vec{v}_{\rm gc} & = \xi \mu v \, \hat{b}_1 + \delta \vec{w} + \frac{\xi^2 m}{q B_0} \hat{b}_0 \times \left[ (\xi \mu v \, \hat{b}_1 + \delta \vec{w}) \cdot \vec{\nabla} \vec{w}_0 \right] + \frac{\xi^2 m}{q B_0} \, \hat{b}_1 \times \left\lbrace \frac{\partial \vec{w}_0}{\partial t} + \left[ \xi \mu v (\hat{b}_0 + \hat{b}_1) + \vec{w}_0 + \delta \vec{w} \right] \cdot \vec{\nabla} \vec{w}_0 \right\rbrace + \\
 & \;\;\;\;\; \frac{\xi^2 m}{q B_0} (\hat{b}_0 + \hat{b}_1) \times \left\lbrace \frac{\partial \delta \vec{w}}{\partial t} + \left[ \xi \mu v (\hat{b}_0 + \hat{b}_1) + \vec{w}_0 + \delta \vec{w} \right] \cdot \vec{\nabla} \delta \vec{w} \right\rbrace + \frac{\xi^3 \mu p}{q B_0} \, \hat{b}_0 \times \left( \delta \vec{w} \cdot \vec{\nabla} \hat{b}_0 \right) + \\
 & \;\;\;\;\; \frac{\xi^3 \mu p}{q B_0} \, \hat{b}_1 \times \left[ \frac{\partial \hat{b}_0}{\partial t} + (\vec{w}_0 + \delta \vec{w}) \cdot \vec{\nabla} \hat{b}_0 \right] + \frac{\xi^3 \mu p}{q B_0} (\hat{b}_0 + \hat{b}_1) \times \left[ \frac{\partial \hat{b}_1}{\partial t} + (\vec{w}_0 + \delta \vec{w}) \cdot \vec{\nabla} \hat{b}_1 \right] + \\
 & \;\;\;\;\; \frac{\xi \mu p}{q B_0} \left[ \frac{\partial \xi}{\partial t} + (\vec{w}_0 + \delta \vec{w}) \cdot \vec{\nabla} \xi \right] (\hat{b}_0 \times \hat{b}_1) + \frac{\xi^2 v p}{q B_0} \left\lbrace \mu^2 (\vec{\nabla} \times \hat{b}_1)_{\perp} + \frac{1 - \mu^2}{2} \left[ (\vec{\nabla} \times \hat{b}_1)_{\parallel} + \frac{\hat{b}_1 \times \vec{\nabla} B_0}{B_0} \right] \right\rbrace + \\
 & \;\;\;\;\; \frac{\xi v p}{q B_0} \left\lbrace \mu^2 \left[ \vec{\nabla} \xi \times (\hat{b}_0 + \hat{b}_1) \right]_{\perp} + \frac{1 - \mu^2}{2} \left[ \vec{\nabla} \xi \times (\hat{b}_0 + \hat{b}_1) \right]_{\parallel} + \frac{1 - \mu^2}{2} \, \vec{\nabla} \xi \times (\hat{b}_0 + \hat{b}_1) \right\rbrace + \\
 & \;\;\;\;\; \frac{\xi^2 v p}{c^2 q B_0} \left\lbrace \mu^2 \left( \delta \vec{w} \times \frac{\partial \hat{b}_0}{\partial t} \right)_{\perp} + \frac{1 - \mu^2}{2} \left[ \left( \delta \vec{w} \times \frac{\partial \hat{b}_0}{\partial t} \right)_{\parallel} +  \frac{\partial B_0}{\partial t} \frac{\hat{b}_0 \times \delta \vec{w}}{B_0} \right] \right\rbrace + \\
 & \;\;\;\;\; \frac{\xi^2 v p}{c^2 q B_0} \left\lbrace \mu^2 \left[ (\vec{w}_0 + \delta \vec{w}) \times \frac{\partial \hat{b}_1}{\partial t} \right]_{\perp} + \frac{1 - \mu^2}{2} \left[ (\vec{w}_0 + \delta \vec{w}) \times \frac{\partial \hat{b}_1}{\partial t} \right]_{\parallel} + \frac{1 - \mu^2}{2} \frac{\partial B_0}{\partial t} \frac{\hat{b}_1 \times (\vec{w}_0 + \delta \vec{w})}{B_0} \right\rbrace + \\
 & \;\;\;\;\; \frac{\xi v p}{c^2 q B_0} \frac{\partial \xi}{\partial t} \left\lbrace \mu^2 \left[ (\vec{w}_0 + \delta \vec{w}) \times (\hat{b}_0 + \hat{b}_1) \right]_{\perp} + \frac{1 - \mu^2}{2} \left[ (\vec{w}_0 + \delta \vec{w}) \times (\hat{b}_0 + \hat{b}_1) \right]_{\parallel} + \frac{1 - \mu^2}{2} (\vec{w}_0 + \delta \vec{w}) \times (\hat{b}_0 + \hat{b}_1) \right\rbrace ,
\end{align*}
where $\hat{b}_0 = \vec{B}_0 / B_0$, $\hat{b}_1 = \delta \vec{B} / B_0$, and $\xi = 1 / \sqrt{1 + \delta B^2 / B_0^2}$. The average velocity is very similar to Eq.~\ref{eq:GyroAvgGCVel} and reduce to its usual definition in the absence of turbulence ($\delta B^2 / B_0^2 = 0$), while the perturbed velocity will vanish in the absence of turbulence ($\delta \vec{B} = \delta \vec{w} = \vec{0}$).

\citet{engelbrechtetal2017} defined the drift velocity, in the isotropic limit and in the absence of electric fields, as a temporal average of the quantity $p v \vec{\nabla} \times (\hat{b}/B) / 3 q$. At this point it should be realised that if the GC velocity is calculated with respect to the total magnetic field, then the gyro- and/or pitch-angle-average will also imply that a suitable average must be taken of the fluctuating fields: 1) the gyro-average for a single particle can be interpreted as a temporal average over a gyration; 2) the gyro-average and the pitch-angle average of a distribution of particles imply an ensemble average because particles of different phases and pitch-angles will sample different parts/realisations of the fluctuations. With an ergodic assumption, the temporal average and ensemble average are equivalent. The assumption that the fields do not vary much over a Larmor radius, also implies, together with the averaging procedure, that turbulence with scales/wavelengths less than the Larmor radius, should be averaged out.

Taking an appropriate average of the above equations then, such that $\langle \delta \vec{B} \rangle = \langle \delta \vec{w} \rangle = \vec{0}$, yields
\begin{align}
\label{eq:AvgGCVelMu}
\vec{V}_{\rm gc} & \approx \sqrt{f_s} \, \mu v \, \hat{b}_0 + \vec{w}_0 +  f_s \left[ \vec{v}_w (\mu) + \sqrt{f_s} \, \vec{v}_t (\mu) + \vec{v}_d (\mu) + \kappa_A^{\rm ws} (\mu) (\vec{\nabla} \times \hat{b}_0)_{\parallel} + \vec{v}_a (\mu) + \frac{\kappa_A^{\rm ws} (\mu)}{c^2} \left( \vec{w}_0 \times \frac{\hat{b}_0}{\partial t} \right)_{\parallel} \right] + \nonumber \\
 & \;\;\;\;\; \frac{3 (1 + \mu^2)}{4} \frac{\kappa_A^{\rm ws}}{c^2} \frac{\partial f_s}{\partial t} (\vec{w}_0 \times  \hat{b}_0) + \frac{3 (1 + \mu^2)}{4} (\vec{\nabla} f_s \times \kappa_A^{\rm ws} \hat{b}_0) ,
\end{align}
where higher order correlations, such as $\langle \hat{b}_1 \times \delta \vec{w} \rangle$, were neglected, $f_s = 1/(1 + \langle \delta B^2 \rangle / B_0^2)$ is identified as the drift suppression factor here, 
\begin{align*}
\vec{v}_w (\mu) & = \frac{3 \kappa_A^{\rm ws}}{v^2} \hat{b}_0 \times \left[ \frac{\partial \vec{w}_0}{\partial t} + (\sqrt{f_s} \mu v \, \hat{b}_0 + \vec{w}_0) \cdot \vec{\nabla} \vec{w}_0 \right] \\
\vec{v}_t (\mu) & = 3 \frac{\mu}{v} \kappa_A^{\rm ws} \hat{b}_0 \times \left( \frac{\partial \hat{b}_0}{\partial t} + \vec{w} \cdot \vec{\nabla} \hat{b}_0 \right) \\
\vec{v}_d (\mu) & = 3 \mu^2 \kappa_A^{\rm ws} (\vec{\nabla} \times \hat{b}_0)_{\perp} + \kappa_A^{\rm ws} (\mu) \frac{\hat{b}_0 \times \vec{\nabla} B_0}{B_0} \\
\vec{v}_a (\mu) & = \frac{1}{c^2} \left[ 3 \mu^2 \kappa_A^{\rm ws} \left( \vec{w}_0 \times \frac{\partial \hat{b}_0}{\partial t} \right)_{\perp} + \kappa_A^{\rm ws} (\mu) \frac{\partial B_0}{\partial t} \frac{\hat{b}_0 \times \vec{w}}{B_0} \right]
\end{align*}
is the drift velocity caused by the temporal and spatial variation of the flow velocity, the drift caused by the temporal and spatial variation of the background magnetic field, the gradient and curvature drift, and the drift velocity caused by the advection of the magnetic field by the flow, respectively, the $(\vec{\nabla} \times \hat{b}_0)_{\parallel}$ and $(\vec{w}_0 \times \hat{b}_0 / \partial t)_{\parallel}$ terms will fall away if the fields are evaluated at the GC's position \citep{rossiolbert1970, burgeretal1985, wijsen2020}, and
\begin{equation}
\label{eq:WeakScatteringLimitMu}
\kappa_A^{\rm ws} (\mu) = \frac{1 - \mu^2}{2} \frac{p v}{q B_0}
\end{equation}
is the pitch-angle dependent weak scattering drift coefficient (notice that this reduces to Eq.~\ref{eq:WeakScatteringLimit} when averaged over pitch-angle).

Notice that the last two terms in Eq.~\ref{eq:AvgGCVelMu} arise due to spatially varying turbulence conditions. Unfortunately, Eq.~\ref{eq:AvgGCVelMu} cannot be simplified further, but it reduces to the isotropic case,
\begin{equation*}
 \left\langle \vec{V}_{\rm gc} \right\rangle_{\mu} = \vec{w}_0 + f_s \frac{3 \kappa_A^{\rm ws}}{v^2} \hat{b}_0 \times \left( \frac{\partial \vec{w}_0}{\partial t} + \vec{w}_0 \cdot \vec{\nabla} \vec{w}_0 \right) + \vec{w}_0 \times \frac{\partial}{\partial t} \left[ f_s \frac{\kappa_A^{\rm ws}}{c^2} \hat{b}_0 \right] + \vec{\nabla} \times (f_s \kappa_A^{\rm ws} \hat{b}_0) ,
\end{equation*}
if averaged over pitch-angle \citep[with some additional terms compared to the usual expressions in][due to the transformation to the flow frame]{rossiolbert1970, burgeretal1985}. It is also interesting to note that a part of the streaming term ($\mu v \, \hat{b}_0$) and the effect of a time dependent magnetic field is zero for the pitch-angle averaged GC drift velocity, as a stationary field is normally explicitly assumed in its derivation \citep[see e.g.][]{burgeretal1985} although it need not be. 
 

\section{Predictions for Solar Energetic Particles}
\label{apndx:Predict}

To fully investigate the implications of the pitch-angle dependent turbulence-reduced drift coefficient derived in Sec.~\ref{sec:Reduce} requires realistic inputs for turbulence quantities such as the magnetic variance, which in turn also influence the behaviour of the perpendicular diffusion coefficient required to evaluate Eq.~\ref{eq:MuDependentFs}. In order to do so in the very inner heliosphere, the present study utilizes values for the various required turbulence quantities calculated by the \citet{zanketal2017} TTM, as solved by \citet{adhikarietal2020}. An in-depth discussion of this model is beyond the scope of this work, but the interested reader is invited to consult, e.g., \citet{zanketal2012,zanketal2017,zanketal2018} and \citet{adhikarietal2017} for more background on this and related models. Exactly the same boundary values and assumptions as discussed by \citet{adhikarietal2020}, reporting model outputs in good agreement with Parker Solar Probe observations of several turbulence quantities at radial distances between $35.55$ and $131.64$ solar radii, are employed here.

\begin{figure*}
\begin{center}
\includegraphics[trim=5mm 13mm 14mm 15mm, clip, width=0.49\textwidth]{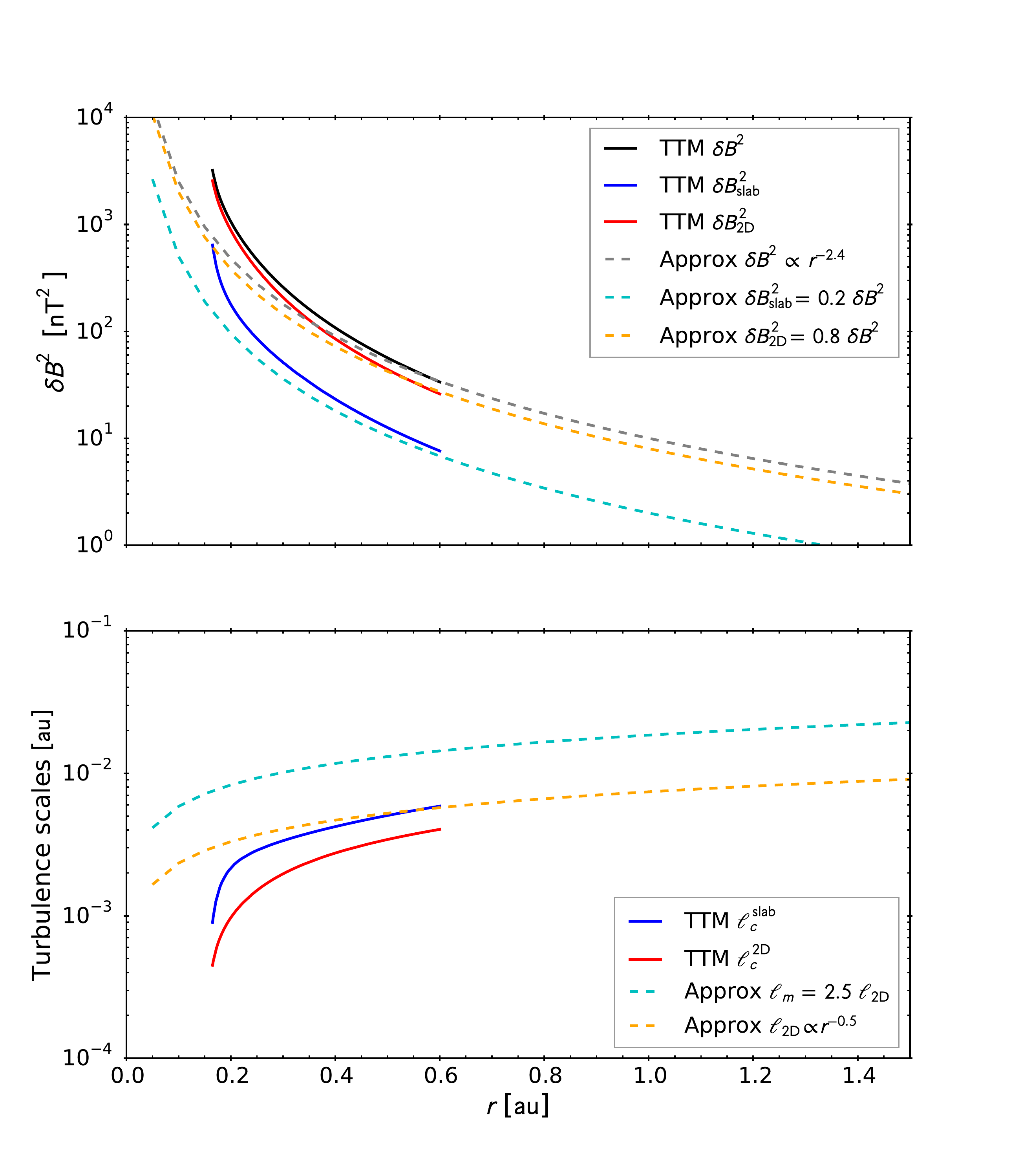}
\includegraphics[trim=5mm 13mm 14mm 15mm, clip, width=0.49\textwidth]{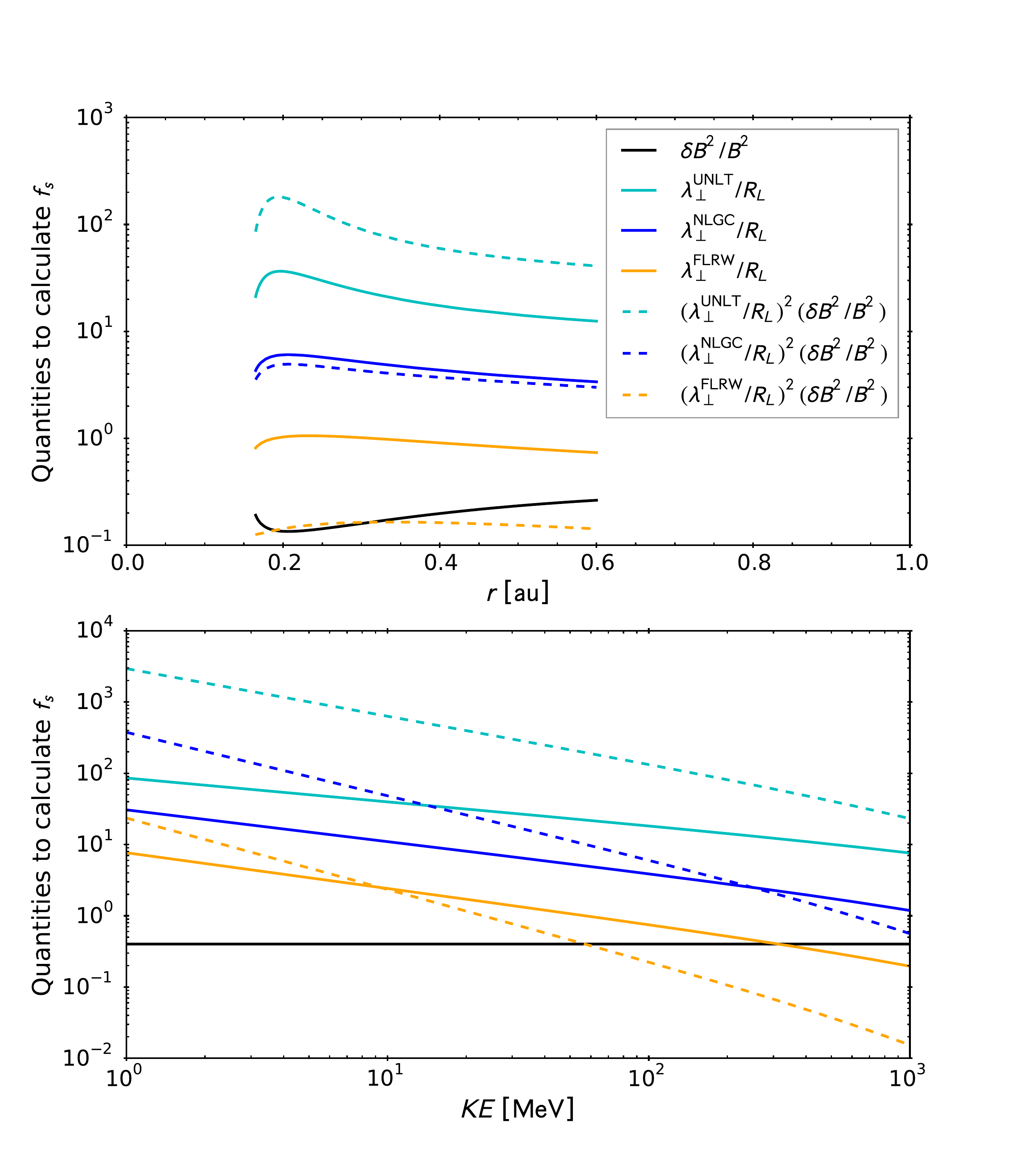}
\caption{\label{fig:TTM}\textit{Top left:} Variance of the magnetic turbulence as a function of radius calculated by the turbulence transport model of \citet{adhikarietal2020} (solid lines), together with some scaling laws used in the outer heliosphere for these quantities normalized to measured values at Earth (dashed lines). \textit{Bottom left:} Correlation lengths of the turbulence as a function of radius calculated by the turbulence transport model of \citet{adhikarietal2020} (blue and red solid lines). The length scales at which the inertial range begins for the slab (dashed cyan) and 2D (dashed orange) turbulence spectra are shown for some assumed scaling laws of these quantities normalized to measured values at Earth. \textit{Right panels:} Various quantities used to calculate the drift reduction factor in Fig.~\ref{fig:Compare} for different theories of perpendicular diffusion, including the parameter $\eta = (\lambda_{\perp}^0 / R_L)^2 (\delta B^2 / B_0^2)$ (dashed lines), as a function of radius (\textit{top right}) for $100~{\rm MeV}$ protons and energy (\textit{bottom right}) for protons at Earth. }
\end{center}
\end{figure*}

Fig.~\ref{fig:TTM} shows magnetic variances (top panel) and correlation scales (bottom panel) yielded by the \citet{adhikarietal2020} model, as function of radial distance. Note that, while this TTM yields several different turbulence quantities, only those directly influencing the MFP expressions discussed in Sec.~\ref{sec:SEPs} are shown in this figure. At the smallest radial distances considered, both slab and 2D variances decrease steeply, before assuming an approximately $r^{-3}$ radial scaling, as reported by \citet{adhikarietal2020}. Furthermore, the values that these quantities assume relative to one another remain roughly in proportion to each other. For the purposes of comparison in Sec.~\ref{sec:SEPs}, the top panel of Fig.~\ref{fig:TTM} also shows simple radial scalings for the total, as well as the slab and 2D variances, as employed in previous studies \cite[see, e.g.,][]{burgeretal2008, molotoetal2018, engelbrecht2019b}, extrapolated to $0.05$~${\rm au}$, and with 2D and slab variances scaled following the 80:20 ratio of the levels of these spectra reported by \citet{bieberetal1994} for observations at $1$~${\rm au}$. A radial scaling of $r^{-2.4}$ is chosen here for the total magnetic variance, normalised to a value of $10$~nT$^2$ at Earth \cite[see, e.g.,][]{smithetal2006, engelbrechtburger2013}, as this roughly corresponds to the radial dependence of variances calculated from Voyager observations beyond $1$~${\rm au}$ \cite[see, e.g.,][]{zanketal1996, smithetal2001}. It should be noted that, due to the differing radial dependencies of the simple scalings and the TTM, resulting variances are considerably different at small radial distances ($<0.25$~${\rm au}$), which would lead to potentially significant differences in transport coefficients.

The slab and 2D correlation scales yielded by the \citet{adhikarietal2020} TTM show, at the smallest radial distances, an increase with radial distance, which becomes less steep beyond $\sim 0.2$~${\rm au}$, corresponding with the decreasing behaviour seen in the magnetic variances. The bottom panel of Fig.~\ref{fig:TTM} also shows simple radial scalings for these correlation lengths, normalised to observational values reported by \citet{weygandetal2011} at Earth, with the radial dependence of $\sqrt{r}$ again chosen to correspond to that displayed by Voyager observations beyond $1$~${\rm au}$ \citep{smithetal2001}. As with the variances, the greatest discrepancies between the TTM outputs and the simple scalings exists at the smallest radial distances considered. However, approaching Earth, the scalings and TTM outputs are similar in value, and display similar radial dependencies. In what is to follow in this appendix, therefore, the outputs yielded by the \citet{adhikarietal2020} TTM are employed as inputs for Eq.~\ref{eq:MuDependentFs} and the various expressions are used to model the perpendicular diffusion coefficients in Sec.~\ref{sec:SEPs}.

\begin{figure*}
\begin{center}
\includegraphics[trim=5mm 2.5mm 14mm 15mm, clip, width=0.975\textwidth]{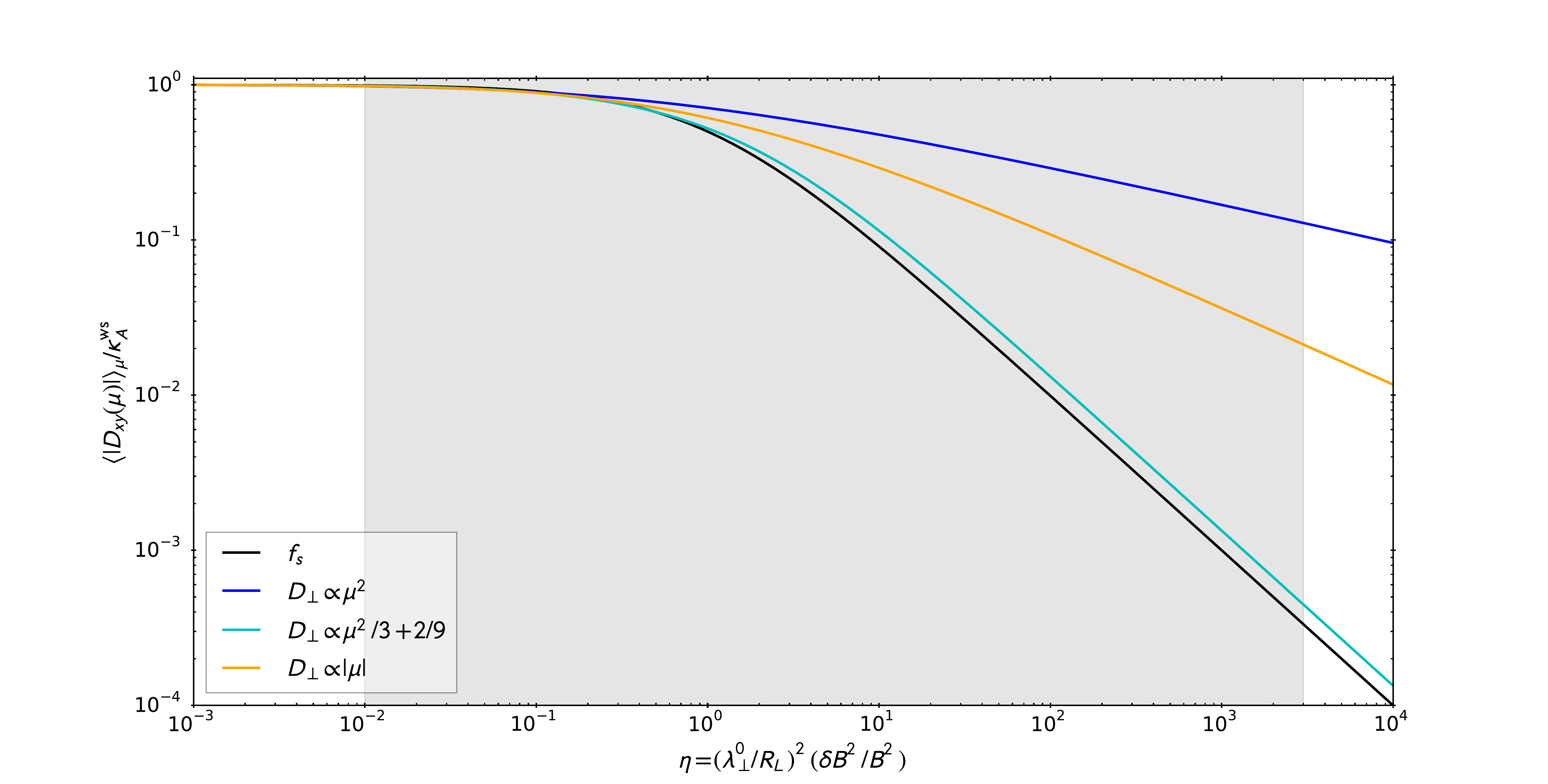}
\caption{\label{fig:ValidityRange}Comparison of the pitch-angle dependent drift coefficient of this work (Eq.~\ref{eq:MuDependentKA}) when averaged over pitch-angle and the isotropic drift coefficient of \citet{engelbrechtetal2017} (Eq.~\ref{eq:IsotropicDriftReduction}; blue). Different functional forms for the pitch-angle dependent perpendicular diffusion coefficient are considered as a function of $\eta = (\lambda_{\perp}^0 / R_L)^2 (\delta B^2 / B_0^2)$. The shaded region indicates the range of $\eta$ values expected from different theories in the inner heliosphere (see the right panel of Fig.~\ref{fig:TTM}).}
\end{center}
\end{figure*}

The top two panels of Fig.~\ref{fig:Compare} show the MFPs corresponding to the various expressions for $D_{\perp}$ discussed in Sec.~\ref{sec:SEPs} along with the parallel MFP and Larmor radius, as function of radial distance for $100$~${\rm MeV}$ protons (left panel), and as function of kinetic energy at Earth (right panel), again for protons, and including the \citet{palmer1982} consensus range values for parallel and perpendicular MFPs at $1$~${\rm au}$. When evaluating these length scales as function of radial distance, the turbulence quantities computed using the \citet{adhikarietal2020} TTM are used. Therefore, plots extend only from $0.165$~${\rm au}$ to $0.6$~${\rm au}$, the radial extent over which those authors solve that model. At Earth, observational values for turbulence quantities reported by \citet{weygandetal2011} and \citet{smithetal2006} are employed. These quantities are known to vary with solar activity \cite[see, e.g.,][]{zhaoetal2018,engelbrechtwolmarans2020}, and, as such, solar minimum values are employed here. In order to discuss the behaviour of these length scales, it is also shown how the ratio of total (slab+2D) variance to HMF magnitude, the ratio of perpendicular MFP to maximal Larmor radius, and the quantity $(\lambda_{\perp}^0/R_L)^2(\delta B^{2}/B_0^2)$, vary as function of radial distance and kinetic energy in the respective middle left and right panels of Fig.~\ref{fig:Compare}. As function of radial distance, all $100$~${\rm MeV}$ proton MFPs, as well as $R_L$, display a relatively flat radial dependence, in line with that of $\delta B^2/B^2$, except at the very smallest radial distances considered, where the steep decrease in correlation scales seen in Fig.~\ref{fig:TTM} strongly influences the behaviour of these MFPs. At $100$~${\rm MeV}$, all perpendicular MFPs remain well below the parallel MFP, with the UNLT expression yielding the largest values for $\lambda_{\perp}^0$ and the FLRW yielding the smallest values, comparable to $R_L$.

At Earth, the parallel MFP remains larger than the \citet{palmer1982} consensus range, due to the fact that solar minimum values for turbulence inputs were chosen \cite[for discussions of this, see][]{engelbrechtburger2013,zhaoetal2018,molotoetal2018}. The NLGC expression for $\lambda_{\perp}^0$ remains closest to the \citet{palmer1982} consensus range for this quantity, with the UNLT result being considerably larger, and displaying a steeper energy dependence due to a stronger dependence on $\lambda_{\parallel}$. The FLRW theory still yields the smallest perpendicular MFP, which is energy independent, and smaller than the Larmor radius above $\sim 60$~${\rm MeV}$. From the middle panels of Fig.~\ref{fig:Compare}, the ratios of $\lambda_{\perp}^0$ to maximal Larmor radius for all scattering theories, except the FLRW, remain larger than unity as function of kinetic energy and radial distance, implying that perpendicular diffusion would play a significant role in the transport of these protons, relative to drift effects. For the FLRW result, this ratio is $\sim 1$ for $100$~${\rm MeV}$ protons, and decreases steadily as function of radial distance. The quantity $(\lambda_{\perp}^0/R_L)^2(\delta B^{2}/B_0^2)$ can also be seen to vary greatly in magnitude, depending on the scattering theory from which $\lambda_{\perp}^0$ is derived, and decreasing as function of kinetic energy, while remaining relatively constant as function of radial distance. Incidentally, the relatively small ($<1$) values for the ratio of $\delta B^2 / B_0^2$ implies that the condition of low turbulence levels imposed by the assumptions implicit in using the \citet{biebermatthaeus1997} approach to deriving the turbulent drift reduction factor is satisfied for the region of the very inner heliosphere considered here.

The bottom two panels of Fig.~\ref{fig:Compare} show the drift reduction factors calculated using the isotropic perpendicular MFPs discussed above as function of radial distance (left panel) and kinetic energy (right panel). As expected from the form of Eq.~\ref{eq:IsotropicDriftReduction}, larger values for $\lambda_{\perp}^0$ lead to smaller values for $f_s$ and hence, a greater reduction in drift effects. As function of radial distance, the $100$~${\rm MeV}$ drift reduction function displays a relatively flat radial dependence, only showing a moderate upturn at the smallest radial distances considered here for the UNLT perpendicular MFP. For the FLRW MFP, almost no drift reduction can be seen at this energy. This is not, however, the case for all energies, as $f_s$ calculated using the FLRW MFP does begin to drop below $1$ at energies below $\sim 100$~${\rm MeV}$. The drift reduction factors for the other MFP expressions considered here also display clear energy dependencies, with the UNLT result remaining well below unity over the entire energy range considered here.

\begin{figure*}
\begin{center}
\includegraphics[trim=40mm 15mm 25mm 15mm, clip, width=0.475\textwidth]{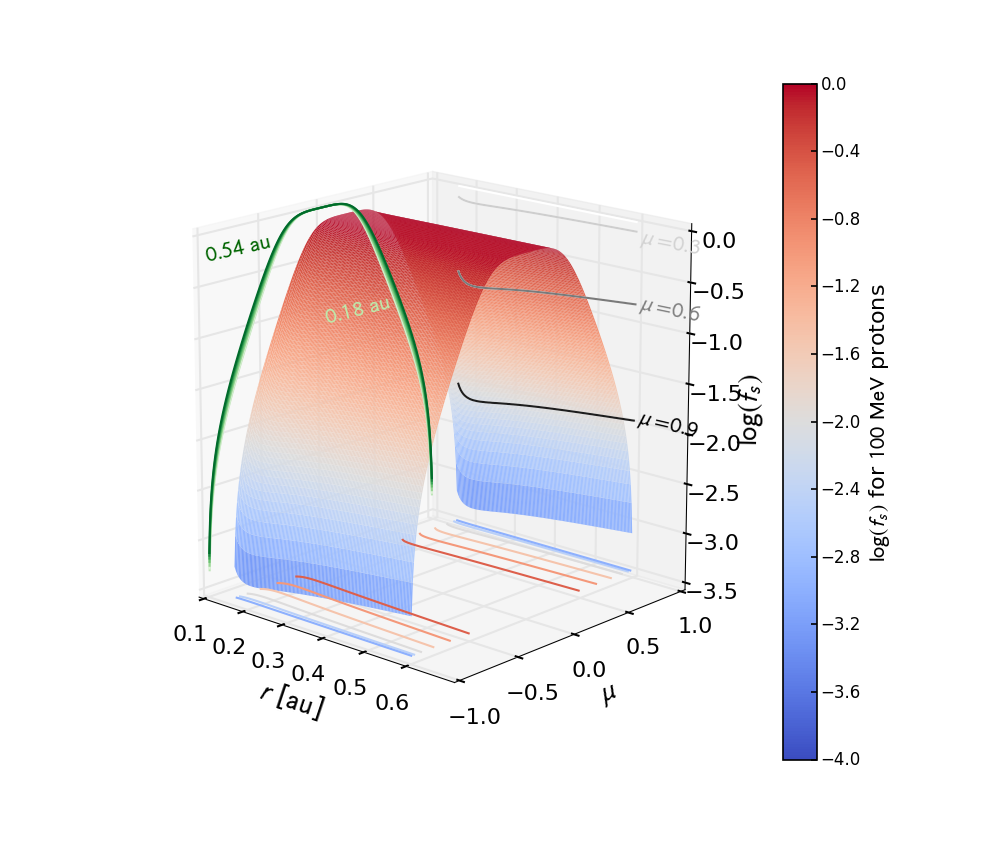}
\includegraphics[trim=40mm 15mm 25mm 15mm, clip, width=0.475\textwidth]{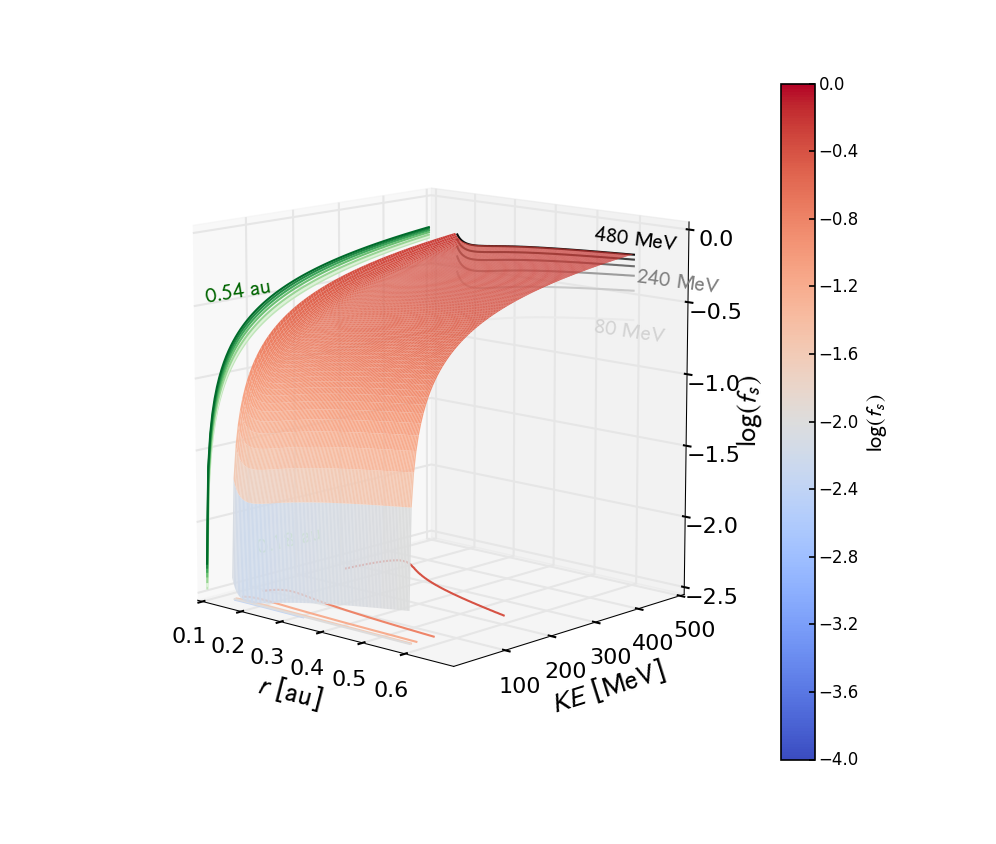}
\caption{\label{fig:Rdependence}\textit{Left:} The pitch-angle dependent drift suppression factor (Eq.~\ref{eq:MuDependentFs}) for $100 \, \mathrm{MeV}$ protons as a function of radius using the quantities in Fig.~\ref{fig:TTM}. \textit{Right:} The isotorpic drift suppression factor (Eq.~\ref{eq:IsotropicDriftReduction}) for protons as a function of radius and kinetic energy using the quantities in Fig.~\ref{fig:TTM}. Perpendicular diffusion is assumed to be described by the non-linear guiding centre theory (Eq.~\ref{eq:NLGC}) in both panels.}
\end{center}
\end{figure*}

The question naturally arises as to when the pitch angle average of Eq.~\ref{eq:MuDependentKA} would correspond to what is expected from Eq.~\ref{eq:IsotropicDriftReduction}. Such a comparison can be done by integrating Eq.~\ref{eq:MuDependentKA}, under the assumption of Eq.~\ref{eq:gmu}, where various pitch-angle dependencies are assumed. Fig.~\ref{fig:ValidityRange} shows the results of these integrations plotted alongside the isotropic result (Eq.~\ref{eq:IsotropicDriftReduction}), as function of $\eta=(\lambda_{\perp}^0/R_L)^2(\delta B^{2}/B_0^2)$. The shaded region on the plot indicates the expected range of values for $\eta$, calculated using the perpendicular MFPs corresponding to the various expressions for $D_{\perp}$ discussed above as function of position and energy (see Fig.~\ref{fig:Compare}), as well as the outputs from the \citet{adhikarietal2020} TTM. For small values of $\eta$, and hence low levels of turbulence, the isotropic and pitch-angle dependent drift reduction factors are very similar. Beyond $\eta \approx 1$, however, results calculated with $g(\mu) = \mu^2$ and $g(\mu) = |\mu|$ (orange and black lines, respectively) begin to deviate considerably from the isotropic case (blue line). It can therefore be concluded that the smaller perpendicular MFPs yielded by the NLGC and FLRW theories (as employed here) will lead to smaller values of $\eta$ that, in turn, implies greater agreement between the pitch-angle averaged approach to the turbulent reduction of drift effects and the isotropic approach. Larger perpendicular MFPs would by implication then lead to a drift reduction factor that is larger than in the isotropic case, but still smaller than one, implying that drift reduction would still occur. Intriguingly, should it be assumed that $g(\mu) = \mu^2/3+2/9$, as motivated by \citet{engelbrecht2019b} based on the numerical test particle simulations of \citet{qinshalchi2014}, the pitch-angle averaged drift reduction factor (cyan line) follows the isotropic result closely. 

Fig.~\ref{fig:EarthDependence} and Fig.~\ref{fig:Rdependence} characterise the pitch-angle, energy, and radial dependence of the turbulence-reduced drift reduction factor. The left panel of Fig.~\ref{fig:EarthDependence} shows the pitch-angle dependence of Eq.~\ref{eq:MuDependentFs} for $100$~${\rm MeV}$ protons and typical turbulence quantities measured at Earth. It is clear that field aligned particles ($\mu = \pm 1$) will have a significant reduction in their drift velocities for all of the different pitch-angle dependencies considered for $D_{\perp} (\mu)$. The behaviour of $f_s(\mu)$ towards $\mu = 0$ depends critically on the theory used for $D_{\perp} (\mu)$. FLRW yields a drift reduction factor mostly $>0.1$, in agreement with Fig.~\ref{fig:Compare}, while $f_s(\mu)$ becomes narrower for NLGC and UNLT. Note that any perpendicular diffusion coefficient which is zero at $\mu = 0$ will have $f_s(0)=1$, and hence particles will experience full drift effects at this pitch-angle. For the form of $g(\mu) = \mu^2/3+2/9$, even particles with $90^{\circ}$ pitch-angles will experience drift reductions. However, there does not seem to be any \textit{a priori} reason that a particle with a $90^{\circ}$ pitch-angle should not experience drift reductions, because it is sampling the transverse turbulence regardless of its pitch-angle. Both the pitch-angle and energy dependence of Eq.~\ref{eq:MuDependentFs} are shown in the right panel of Fig.~\ref{fig:EarthDependence} at Earth for NLGC. The drift reduction factor $f_s$ becomes more pronounced at low energies, for a fixed pitch-angle (except at $\mu = 0$), causing $f_s(\mu)$ to become narrower in $\mu$ at low energies.

The left panel of Fig.~\ref{fig:Rdependence} shows the pitch-angle and radial dependence of the drift suppression factor for $100$~${\rm MeV}$ protons assuming the NLGC result for the perpendicular diffusion (Eq.~\ref{eq:NLGC}). The shape of $f_s(\mu)$ does not change much with radius, although an upturn can be seen in the very inner heliosphere, in agreement with what can be seen in Fig.~\ref{fig:Compare}. The isotropic drift suppression factor's energy and radial dependence is shown in the right panel of Fig.~\ref{fig:Rdependence}, again for NLGC. The radial dependence is again weaker than the energy dependence, but shows the upturn at small radial distances. It is clear that low energy particles will experience much more drift reduction than high energy particles. Similar plots can, of course, also be made for FLRW or UNLT, but the qualitative behaviour of the drift reduction factor will still be the same: field aligned particles will experience more drift reductions, low energy particles will experience less drifts, and drift reduction factors will have a weak radial dependence.


\end{document}